\documentclass[twocolumn]{aastex701}

\newcommand{\ciii}{C~III] $\lambda 1908$}
\newcommand{\civ}{C~IV $\lambda 1549$}

\newcommand{\niii}{N~III] $\lambda 1750$}
\newcommand{\niv}{N~IV] $\lambda 1488$}

\newcommand{\oiiiboth}{[O~III] $\lambda\lambda 4959, 5007$}

\usepackage{graphicx}
\usepackage{subcaption}
\usepackage[dvipsnames]{xcolor}

\begin{document}

\title{JWST CAPERS: Spectroscopic evaluation of very high redshift galaxy candidates}

\author[0000-0002-7622-0208]{Callum T. Donnan}
\affiliation{NSF's National Optical-Infrared Astronomy Research Laboratory, 950 N. Cherry Ave., Tucson, AZ 85719, USA}
\email[show]{callum.donnan@noirlab.edu}

\author[0000-0001-5414-5131]{Mark Dickinson}
\affiliation{NSF's National Optical-Infrared Astronomy Research Laboratory, 950 N. Cherry Ave., Tucson, AZ 85719, USA}
\email{mark.dickinson@noirlab.edu}

\author[0000-0002-8951-4408]{Lorenzo Napolitano}
\affiliation{INAF - Observatorio Astronomico di Roma, Via Frascati 33, 00078 Monte Porzio Catone, Rome, Italy}
\email{lorenzo.napolitano@inaf.it}

\author[0000-0002-7959-8783]{Pablo Arrabal Haro}
\affiliation{Center for Space Sciences and Technology, UMBC, 5523 Research Park Dr, Baltimore, MD 21228 USA }
\affiliation{Astrophysics Science Division, NASA Goddard Space Flight Center, 8800 Greenbelt Rd, Greenbelt, MD 20771, USA}
\email{pablo.arrabalharo@nasa.gov}

\author[0000-0003-1282-7454]{Anthony J. Taylor}
\email{anthony.taylor@austin.utexas.edu}
\affiliation{Department of Astronomy, The University of Texas at Austin, Austin, TX, USA}
\affiliation{Cosmic Frontier Center, The University of Texas at Austin, Austin, TX, USA}

\author[0000-0003-3596-8794]{Hollis B. Akins}
\email{hollis.akins@gmail.com}
\affiliation{Department of Astrophysical Sciences, Princeton University, 4 Ivy Lane, Princeton, NJ 08544, USA}
\altaffiliation{NASA Einstein Fellow}
\email{hollis.akins@gmail.com}

\author[0000-0003-0629-8074]{Ryan Begley}
\affiliation{Armagh Observatory and Planetarium, College Hill, Armagh, BT61 9DG, N. Ireland, UK}
\affiliation{Institute for Astronomy, University of Edinburgh, Royal Observatory, Edinburgh, EH9 3HJ, UK}
\email{ryan.begley@armagh.ac.uk}

\author[0000-0001-8863-2472]{Davide Bevacqua}
\affiliation{INAF - Observatorio Astronomico di Roma, Via Frascati 33, 00078 Monte Porzio Catone, Rome, Italy}
\email{davide.bevacqua@inaf.it}  

\author[0000-0002-1482-5818]{Adam C. Carnall}
\affiliation{Institute for Astronomy, University of Edinburgh, Royal Observatory, Edinburgh, EH9 3HJ, UK}
\email{adamc@roe.ac.uk}

\author[0000-0001-9875-8263]{Marco Castellano}
\affiliation{INAF - Osservatorio Astronomico di Roma, Via Frascati 33, 00078 Monte Porzio Catone, Rome, Italy}
\email{marco.castellano@inaf.it}

\author[0000-0002-3736-476X]{Fergus Cullen}
\affiliation{Institute for Astronomy, University of Edinburgh, Royal Observatory, Edinburgh, EH9 3HJ, UK}
\email{fergus.cullen@ed.ac.uk}

\author[0000-0002-1404-5950]{James S. Dunlop}
\affiliation{Institute for Astronomy, University of Edinburgh, Royal Observatory, Edinburgh, EH9 3HJ, UK}
\email{james.dunlop@ed.ac.uk}

\author[0000-0001-8519-1130]{Steven L. Finkelstein}
\affiliation{Department of Astronomy, The University of Texas at Austin, Austin, TX, USA}
\affiliation{Cosmic Frontier Center, The University of Texas at Austin, Austin, TX, USA}
\email{stevenf@astro.as.utexas.edu}

\author[0009-0000-0272-5468]{Carter Flayhart}
\affiliation{Laboratory for Multiwavelength Astrophysics, School of Physics and Astronomy, Rochester Institute of Technology, 84 Lomb Memorial Drive, Rochester, NY 14623, USA}
\email{cbf2388@rit.edu}

\author[0000-0003-3820-2823]{Adriano Fontana}
\affiliation{INAF - Osservatorio Astronomico di Roma, via di Frascati 33, 00078 Monte Porzio Catone, Italy}
\email{adriano.fontana@inaf.it}

\author[0000-0002-7831-8751]{Mauro Giavalisco}
\affiliation{University of Massachusetts Amherst, 710 North Pleasant Street, Amherst, MA 01003-9305, USA}
\email{mauro@umass.edu}

\author[0000-0000-0000-0000]{Michaela Hirschmann}
\affiliation{Institute of Physics, Lab for galaxy evolution and spectral modelling, EPFL, Observatory of Sauverny, Chemin Pegasi 51, 1290 Versoix, Switzerland}
\email{michaela.hirschmann@epfl.ch}

\author[0000-0003-1581-7825]{Ray A. Lucas}
\affiliation{Space Telescope Science Institute, 3700 San Martin Drive, Baltimore, MD 21218, USA}
\email{lucas@stsci.edu} 

\author[0000-0003-4368-3326]{Derek J. McLeod}
\affiliation{Institute for Astronomy, University of Edinburgh, Royal Observatory, Edinburgh, EH9 3HJ, UK}
\email{derek.mcleod@ed.ac.uk}

\author[0009-0005-9742-2318]{Ross J. McLure}
\affiliation{Institute for Astronomy, University of Edinburgh, Royal Observatory, Edinburgh, EH9 3HJ, UK}
\email{rmclure@ed.ac.uk}

\author[0000-0001-7503-8482]{Casey Papovich}
\affiliation{Department of Physics and Astronomy, Texas A\&M University, College Station, TX, 77843-4242 USA}
\affiliation{George P.\ and Cynthia Woods Mitchell Institute for Fundamental Physics and Astronomy, Texas A\&M University, College Station, TX, 77843-4242 USA}
\email{papovich@tamu.edu}

\author[0000-0001-8940-6768]{Laura Pentericci}
\affiliation{INAF - Observatorio Astronomico di Roma, Via Frascati 33, 00078 Monte Porzio Catone, Rome, Italy}
\email{laura.pentericci@inaf.it }

\author[0000-0002-0939-9156]{Borja Pérez-Díaz}
\affiliation{INAF - Observatorio Astronomico di Roma, Via Frascati 33, 00078 Monte Porzio Catone, Rome, Italy}
\email{borja.perezdiaz@inaf.it}

\author[0000-0002-0827-9769]{Thomas M. Stanton}
\affiliation{Institute for Astronomy, University of Edinburgh, Royal Observatory, Edinburgh, EH9 3HJ, UK}
\email{T.Stanton@ed.ac.uk}

\author[0000-0003-3466-035X]{{L. Y. Aaron} {Yung}}
\affiliation{Space Telescope Science Institute, 3700 San Martin Drive, Baltimore, MD 21218, USA}
\affiliation{Institute of Astronomy and Department of Physics, National Tsing Hua University, Hsinchu 30013, Taiwan}
\email{l.y.aaronyung@gmail.com}

\begin{abstract}

\textit{JWST}/NIRCam imaging has enabled efficient selection of a large number of galaxy candidates at very high redshifts. Accurate and robust selection is crucial for statistical analyses, such as measurement of the ultraviolet luminosity function (UV LF). A variety of selection methods and criteria have been proposed in the literature. 
We evaluate the selection methodology of galaxy candidates at $z\geq8.5$ using 183 candidates with secure spectroscopic redshifts obtained from the CAPERS \textit{JWST}/NIRSpec PRISM program. The success of high-$z$ confirmation is strongly correlated with the photometric signal-to-noise ratio (SNR) in the detection filter, the strength of the Lyman-$\alpha$ break, and the $\Delta \chi^2$ between the best-fitting low-redshift and high-redshift spectral energy distribution model. However, the most efficient selection method is the use of a UV LF prior on the inferred $p(z)$ of a source, raising the success rate from $\simeq 35\%$ to $\simeq 77\%$ while achieving greater completeness than conservative SNR and $\Delta \chi^2$ cuts. Out of the 183 candidates we spectroscopically confirm 48 sources at $z\geq8.5$ via detections of emission lines and/or the Lyman-$\alpha$ break (particularly important for those with no emission line detection). We report the highest-redshift source in CAPERS as UDS-111893 at $z_{\mathrm{spec}}=12.98$, with no sources confirmed at $z>13$. This work provides spectroscopic quantification of \textit{JWST}/NIRCam photometric selection at $z\geq8.5$, 
and demonstrates the effectiveness of a UV LF prior for high-redshift photometric selection.

\end{abstract}



\section{Introduction}

Understanding the formation of the first galaxies is vital for building a complete picture of galaxy evolution and cosmology. The crucial first step in building this understanding is accurate selection of these galaxies from near-infrared imaging. The launch of \textit{JWST} has allowed the robust detection of galaxies at $z>9$ for the first time with sensitive imaging from $\simeq2-5\mu \mathrm{m}$ with the NIRCam instrument. There is now an abundance of galaxy candidates at $z>9$ \citep[e.g.,][]{castellano2022,naidu2022,donnan2023b, donnan2023a,adams2022, finkelstein2022c,harikane2023a,franco2024} identified from a number of now-complete \textit{JWST}/NIRCam surveys. These galaxy candidates have been used to measure the evolution of the UV luminosity function (LF) out to $z\sim14$ showing a more shallow evolution \citep[e.g.,][]{finkelstein2024, perezgonzalez2023,2025arXiv250315594P, mcleod2024, adams2023, donnan2024,whitler2025}  than predicted by many pre-\textit{JWST} theoretical models \citep{mason2015,tacchella2018,yung2019} but in agreement with some pre-\textit{JWST}  observations \citep[e.g.][]{mcleod2016}. Recent work covering large areas with \textit{JWST} ($\sim 0.5\ \mathrm{deg}^2$) suggests that this shallow evolution may not continue at $z>13$, with a more rapid decline in the number density of galaxies at $z>13$ \citep{weibel2026,kreilgaard2026,mcleod2026}.

In addition to robust photometric selection of very high redshift (``VHZ'', $z\geq8.5)$ galaxy candidates, \textit{JWST}  also enables efficient redshift verification using NIRSpec  multi-object spectroscopy. NIRSpec has confirmed the redshifts of many galaxies at $z>9$ \citep[e.g.,][]{curtislake2022, bunker2023, arrabalharo2023, carniani2024, castellano2024,napolitano2024, napolitano2025, kokorev2025,donnan2025b,donnan2026}, with the highest-redshift, spectroscopically-confirmed galaxy currently at $z_{\mathrm{spec}}=14.44$ \citep{naidu2025}. NIRSpec has also allowed the characterization of the physical properties at $z>9$ in detail for the first time. A number of these galaxies display strong rest-frame UV emission lines (e.g. \niii, \niv) which imply elevated nitrogen abundances inconsistent with many chemical evolution models \citep[e.g.,][]{cameron2023,castellano2024,naidu2025}. In contrast a number of these galaxies have weak or a complete absence of UV emission lines \citep{curtislake2022,arrabalharo2023,donnan2026} with a potential connection to physical size with the strong-line emitters appearing more compact \citep{harikane2024}. A potential explanation for these seemingly two separate populations is that they represent different stages of an evolutionary path with a stochastic star-formation history \citep{cole2025,robertsborsani2026,mcclymont2026}. 


However, there is a variety of selection methods and criteria used to select galaxies at $z>9$. This is particularly apparent at $z>13$ where some methods find an abundance of galaxy candidates at $z>13$ \citep[e.g.][]{kokorev2025b,castellano2025,perezgonzalez2025} whereas more conservative criteria yield a dearth of candidates at these extreme redshifts leading to the implication of a more rapid evolution of the UV LF at $z>13$ \citep{weibel2026,kreilgaard2026,mcleod2026}. 

Spectroscopic verification with NIRSpec is the ideal path to evaluating the most effective method of selecting galaxies at $z>9$. However, although NIRSpec has successfully confirmed the redshifts of many galaxies at these redshifts, these have often resulted from small programs where the targets are highly biased to the most robust candidates. To resolve this requires a NIRSpec program that can target the robust candidates at $z>9$ in a given field, but is
of sufficient size that it can also target a number of less robust $z>9$ candidates. This therefore builds a sample of
$z>9$ targets selected with a range of photometric robustness allowing the success of different selection criteria to
be quantified. The CANDELS-Area Prism Epoch of Reionization Survey (CAPERS; GO-6368, PI: M.\ Dickinson) addresses this aim with extensive NIRSpec multi-object spectroscopy of VHZ galaxy candidates over three fields
imaged with \textit{JWST}/NIRCam. 

In this work we present the sample of galaxy candidates at $z_{\mathrm{phot}}\geq8.5$ targeted by the CAPERS program and present their spectroscopic results. We use this to evaluate the success of different selection criteria quantifying their success using this large sample of $z\geq8.5$ targets for the first time. Finally, we present the full sample of galaxies confirmed to be at $z_{\mathrm{spec}}\geq8.5$ in the CAPERS survey and discuss their observational properties.

This paper is structured as follows. In Section~\ref{sec:data} we introduce the photometric and spectroscopic data. In Section~\ref{sec:selection_techniques} we quantify a number of commonly used photometric selection criteria. In Section~\ref{sec:z_failure} we discuss the nature of those sources that are low-redshift interlopers and in Section~\ref{sec:conf_VHZ} we present those sources in CAPERS which are confirmed to be at $z\geq8.5$. Finally, in Section~\ref{sec:conclusion} we summarize our conclusions. Throughout we use magnitudes in the AB system \citep{oke1974,oke1983}, a \citet{Kroupa2001} initial mass function, and assume a standard cosmological model with $H_0=70$ km s$^{-1}$ Mpc$^{-1}$, $\Omega_m=0.3$ and $\Omega_{\Lambda}=0.7$.  

\section{Data}
\label{sec:data}

\subsection{Imaging, Photometry, and Photometric Redshifts}
\label{sec:photometry}
CAPERS targets galaxies across three different legacy extragalactic fields - COSMOS, UDS and EGS. These fields have rest-frame optical imaging in the \textit{HST}/ACS F435W, F606W, F814W filters  \citep{davis2007,koekemoer2007,koekemoer2011,grogin2011}, and have also been extensively imaged with \textit{JWST}/NIRCam F090W, F115W, F150W, F200W, F277W, F356W, F410M and F444W. The NIRCam imaging used for target selection was primarily obtained by the Cosmic
Evolution Early Release Science \citep[CEERS;][]{finkelstein2025} program, with additional imaging from GO-2234 (PI: E.\ Ba\~{n}ados) and DDT-2750 (PI: P.\ Arrabal-Haro). The NIRCam imaging in UDS and COSMOS was obtained by the Public Release IMaging for Extragalactic Research survey (PRIMER; Dunlop et al.\ in prep.).

The primary photometric catalogs used for CAPERS observation planning are early versions of the UNICORN catalogs (S.\ Finkelstein et al., in prep.), which follow a similar methodology for source detection and photometry to that described in \citet{finkelstein2024}. The UNICORN catalogs include photometric redshifts computed using \textsc{eazy} \citep{brammer2008}. As will be discussed further in \S\ref{sec:vhz_selection}, VHZ target prioritization incorporates additional catalogs and photometric redshifts computed with different methods, but we will refer to the UNICORN catalogs and photometric redshifts as the ``reference'' or ``fiducial'' sample. 

\subsection{Spectroscopy}
\label{sec:spectroscopy}
CAPERS is a \textit{JWST}/NIRSpec Cycle 3 Treasury Program which uses the PRISM mode which provides low resolution ($R \approx 30$ to 300) spectra over the wavelength range $0.6-5.3 \mu \mathrm{m}$.
The CAPERS observations used for the present analysis consist of 21 different NIRSpec pointings across the COSMOS, UDS and EGS fields (7 pointings per field). Each pointing typically comprises three Multi-Shutter Assembly \citep[MSA;][]{ferruit2022} configurations, except for UDS pointings P2, P3 and P5, which have just two configurations each. Each MSA configuration has an exposure time of 5689.67~s (1.58 hr). A number of the highest-priority targets obtain observations over multiple exposures leading to a maximum exposure time of  17069~s (4.74 hr) for some targets, including many of the VHZ candidates. 

The NIRSpec data in UDS, COSMOS and EGS were uniformly reduced using version 1.20.2 of the NIRSpec reduction pipeline with CRDS version \texttt{1464.pmap}. This process follows the general method in \citet{arrabalharo2023b} with some modifications. The \texttt{clean\_flicker\_noise} option is used to remove the $1/f$ noise in the spectra in the \texttt{calwebb\_detector1} stage. In
\texttt{calwebb\_spec2} we also use a modified flat-field and after
running the \texttt{calwebb\_spec3} stage we obtain a reduced 2D spectrum. Finally, the 1D spectra are produced using an optimal extraction \citep{horne1986} whenever the galaxy continuum is detected in the spectrum with sufficient signal-to-noise ratio to define a weighting profile.

\subsection{Spectroscopic redshift measurement}
\label{sec:z_spec_measurement}
We measure spectroscopic redshifts of high redshift candidates in CAPERS following the approach of \citet{castellano2026,napolitano2025}; Napolitano et al.\ in prep. For sources with emission lines detected at $\mathrm{SNR} >3$, we derive an SNR-weighted centroid for each available significant line. The spectroscopic redshift is then computed as the inverse-variance weighted average of the redshifts inferred from the individual lines.

In the case of sources with likely $z > 6$ but without detected emission lines, we measure the redshift of the source by modeling the Lyman-$\alpha$ break. We implement a model of Lyman-$\alpha$ break where there is complete IGM attenuation at $\lambda / \mu \mathrm{m}<0.1216(1+z)$ with continuum modeled by a power law of the form $f_{\lambda}\propto \lambda^{\beta}$ redward of the Lyman-$\alpha$ break. We introduce damping on the Lyman-$\alpha$ break from the neutral hydrogen at $z>6$ following the method of \citep{miralda-escude1998} where the IGM neutral hydrogen fraction is given by $X_{\mathrm{HI}}$. In Appendix~\ref{sec:apdx_z_break} we test the need to introduce additional damping from local hydrogen using the prescription from \citep{heintz2024}. However, we find that the highest-redshift accuracy is achieved with an IGM only model. Finally, the model is convolved with the wavelength dependent spectral resolution of NIRSpec PRISM and fitted to the CAPERS spectra using dynamic nested sampling with \textsc{dynesty} \citep{speagle2020}. We assume the following uniform priors on the free parameters: $6.5\leq z\leq20$, $-4\leq \beta\leq4$, $0\leq X_{\mathrm{HI}}\leq1$ and restrict the data to where $\lambda_{\mathrm{obs}}<3600 (1+z)\, \mathrm{\AA}$, where $z$ is the initial estimated redshift based on the position of the Lyman-$\alpha$ break.

Finally, all spectroscopic redshift measurement are assigned a quality flag ($z_{\mathrm{q}}$) following the classification method of \citet{pentericci2018} where the highest quality flags are $z_{\mathrm{q}} = 3$, 4 and 9. $z_{\mathrm{q}} = 3$ and 4 indicate secure confirmations supported by several well-detected spectral features. $z_{\mathrm{q}} = 9$ indicates that there is a single, well-detected feature, whose identification is supported by other evidence. For the redshift range discussed here, that feature is generally a strong continuum break, and additional evidence, such the photometric redshift or significant blueward flux, is considered to evaluate whether it is likely to be the Lyman-$\alpha$ break.

\subsection{Selection of $z>8.5$ galaxy targets}
\label{sec:vhz_selection}
The primary targets of the CAPERS program are defined as those with a photometric redshift of $z_{\mathrm{phot}}\gtrsim8.5$ which we label as ``very high redshift'' (VHZ) candidates. It is frequently the case that photometric redshift probability distributions ($P[z_{\mathrm{phot}}]$) have two or more peaks at very different redshifts.  This may lead to confusion between objects that genuinely have very high redshifts, and foreground interlopers. Different studies have adopted different criteria to define ``robust'' high redshift candidates, such as thresholds in $\Delta\chi^2$, the difference between the goodness-of-fit parameters models fixed to the high- and low-redshift solutions (see \S\ref{sec:deltachi2}). Other examples will be discussed later in this paper.

CAPERS deliberately casts a wide net for VHZ candidates in order to explore the boundaries of high redshift selection criteria and to better understand the nature of foreground contaminants. The overall VHZ sample analyzed here is not limited to the most ``robust'' selection criteria, and it is expected (and intentional) that there will be a significant number of interlopers.  We collected VHZ candidates using a variety of selection criteria, contributed by several individuals or groups both within the CAPERS investigator team and contributed by members of the wider community, sometimes using different reductions of the imaging data and/or different photometric catalogs. In some cases, we used samples in which robustness criteria such as $\Delta\chi^2$ were deliberately relaxed.  The CAPERS VHZ target list is the superset of galaxies that were assigned a likely redshift $z > 8.5$ in at least one of the contributed subsamples. A more complete discussion of the VHZ target samples and their prioritization will be presented in a forthcoming CAPERS survey overview paper. 

Throughout this work we quote two quantities. First, we define a successful spectroscopic confirmation as $|dz|<0.15$ where $dz = (z_{\mathrm{phot}} - z_{\mathrm{spec}})/(1+z_{\mathrm{spec}})$. This is commonly used to define ``catastrophic redshift failure'' in the case where $|dz|>0.15$ \citep[e.g.,][]{mcleod2021} whereas here we use the opposite to define a redshift success. Here we define $f_{\mathrm{success}}$ as the number of sources with $|dz|<0.15$ divided by the relevant total sample size:
\begin{equation}
    f_{\mathrm{success}}(X) = \frac{N(z_{\mathrm{phot}}\geq8.5,X,|dz|<0.15)}{N(z_{\mathrm{phot}}\geq8.5,X)}
\end{equation}
where $X$ is a given selection criteria (e.g., minimum $\mathrm{SNR}$).
The completeness, $f_{\mathrm{complete}}$, is the fraction of CAPERS VHZ targets with $z_{\mathrm{spec}}\geq8.5$ that are recovered with $|dz|<0.15$ divided by relevant total sample size: 
\begin{equation}
    f_{\mathrm{complete}}(X) = \frac{N(z_{\mathrm{spec}}\geq8.5,X,|dz|<0.15)}{N(z_{\mathrm{spec}}\geq8.5)}
\end{equation}
We note that the latter is completeness relative to the CAPERS VHZ sample rather than to the true underlying population, since galaxies at $z\geq8.5$ that were not selected by any contributing catalog did not enter the sample.

CAPERS observed a total of 327 VHZ targets (149 in COSMOS, 82 in UDS and 96 in EGS). From the observed VHZ candidates, we define successful spectroscopic redshift measurements as those with a redshift quality flag $ z_{\mathrm{q}} =3, 4$ or $9$ yielding a sample of 183. Of these 183 sources, 48 have $z_{\mathrm{spec}}\geq8.5$ with therefore 135 sources at $z_{\mathrm{spec}}<8.5$. The remaining 144 VHZ sources fail to yield a secure redshift measurement due to a number of factors such as large wavelength gaps, strong contamination from neighboring sources or simply insufficient signal-to-noise.

\begin{figure*}[ht!]
\centering
\includegraphics[width=\linewidth]
{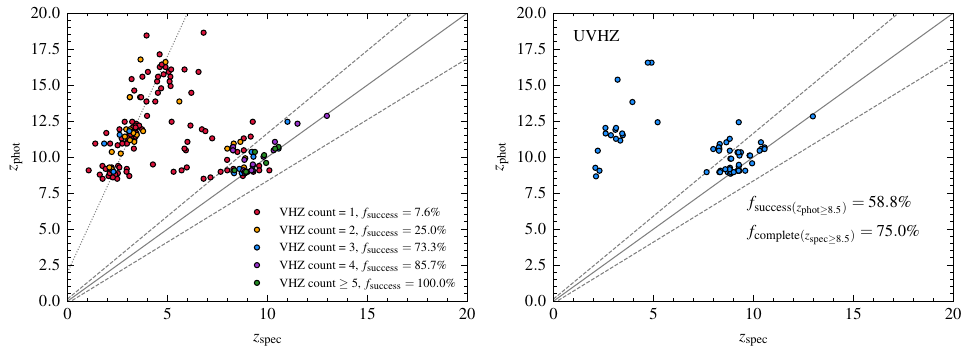}
\caption{\textit{Left}: A comparison of photometric redshifts of the 183 CAPERS VHZ targets to the measured spectroscopic redshifts split by the number of times the target was selected by different VHZ subsamples. For this analysis, we force the reference photometric redshifts to match the highest peak probability $P(z_{\mathrm{phot}})$ at $z > 8.5$ using \textsc{eazy} with the templates from \citet{larson2023a}. The inset success rate is defined as the percentage of sources where $|dz|<0.15$, where $dz = (z_{\mathrm{phot}} - z_{\mathrm{spec}})/(1+z_{\mathrm{spec}})$. Sources that fall in the region outlined by the dashed gray lines are those with $|dz|<0.15$. The dotted line is given by $\frac{3646}{1216}(1+z_{\mathrm{spec}})  -1$ illustrating that their true redshift is where the observed photometric break is given by a Balmer break rather than a Lyman-$\alpha$ break. \textit{Right}: A comparison of the photometric redshifts of one ``robustly-selected'' subsample (UVHZ; following method of \citealp{finkelstein2024}) with the success rate and completeness rate of this specific sample.}
\label{fig:vhz_zz}
\end{figure*}

In Fig.~\ref{fig:vhz_zz} we show a comparison of the photometric redshifts to the spectroscopic redshifts obtained with CAPERS. 
The community-sourced nature of the VHZ targets in CAPERS means that a given VHZ candidate may have $z_{\mathrm{phot}} > 8.5$ in some photometric redshift catalogs and lower redshift in others. The classification of a source as a VHZ candidate simply requires a single catalog to find it preferred to be at $z\geq8.5$. 
Although the reference photometric redshift measurements from UNICORN (\S\ref{sec:photometry}) place some VHZ candidates at $z_{\mathrm{phot}}<8.5$, all candidates also have $P(z_{\mathrm{phot}})$ peaks at $z_{\mathrm{phot}} > 8.5$, and for the left-hand panel of Fig.~\ref{fig:vhz_zz} (only) we force the reference photometric redshifts to match the high-$z$ peak.

We find that the rate of successful spectroscopic confirmation rises with increasing ``VHZ count'', defined as the number of contributed subsamples that selected the target as a VHZ candidate. The success rate rises from $\sim7\%$ to $\sim100\%$ from VHZ count $=1$ to 5. Out of the total sample of 183 VHZ targets, there are 119 with VHZ count $=1$, 28 with VHZ count $=2$, 15 with VHZ count $=3$, 14 with VHZ count $=4$ and 7 with VHZ count $\geq5$. The maximum VHZ count for any observed target is 7, for EGS-25297 \citep{donnan2025b}. This shows a strong correlation between successful VHZ spectroscopic confirmation and the number of different catalogs that selected the candidate:  consensus is a good predictor of success. 

For many candidates spectroscopically confirmed as low-$z$ interlopers, their observed photometric break is in reality a Balmer break rather than a Lyman-$\alpha$ break. This is demonstrated by the dotted line in Fig.~\ref{fig:vhz_zz}.

In the right-hand panel of Fig.~\ref{fig:vhz_zz}, we show the comparison between photometric and spectroscopic redshifts for one of the more ``robust'' samples used in target selection. This ``Unicorn VHZ'' (UVHZ) sample implements the selection criteria of \citet{finkelstein2024}, including several parameters discussed below in \S\ref{sec:selection_techniques}, such as thresholds in detection S/N and $\Delta\chi^2$, strict limits on S/N blueward of the Lyman-$\alpha$ break, and requirements on the best-fitting $z_{\mathrm{phot}}$ and the integrated $P(z_{\mathrm{phot}})$ at high $z_{\mathrm{phot}}$. The success and completeness rates for this UVHZ subsample are higher than those for the VHZ sample overall. However, there are still spectroscopically confirmed interlopers, and not all galaxies with confirmed CAPERS $z_{\mathrm{spec}} > 8.5$ were selected by the UVHZ criteria.

In both panels, there are a sizable numbers of ``near-misses'':  objects with $z_{\mathrm{phot}}$ slightly higher than the $dz = +0.15$ threshold we use to delimit spectroscopic ``success''.  While their spectroscopic redshifts are lower than the photometric predictions, these ``failures'' are still reionization-era galaxies with $6.8 \lesssim z \lesssim 9.2$ (and $z > 8$ for UVHZ). If we adopt a more generous threshold of success, with $dz < 0.35$, the overall UVHZ $f_{\mathrm{success}}$ increases from 58\% to 73\%, and $f_{\mathrm{complete}}$ increases from 75\% to 83\%. These ``near-misses'' will be discussed further in \S\ref{sec:near_misses}.

\begin{figure}[ht!]
\centering
\includegraphics[width=\columnwidth]
{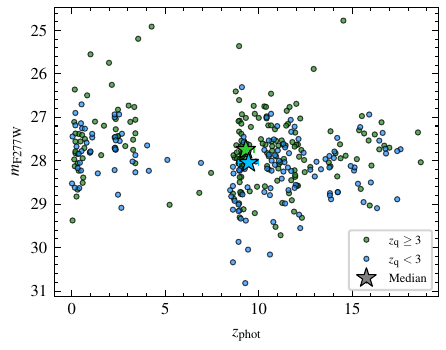}
\caption{The photometric redshift (using \textsc{eazy} with the templates from \citet{larson2023a}) and apparent magnitude in the F277W filter for the total VHZ target sample (N=327) separated into those with secure redshift measurements (green data-points; $z_{\mathrm{q}} \geq3$) and those where we fail to measure a secure spectroscopic redshift (blue data-points; $z_{\mathrm{q}} < 3$). The starred data-points indicate the median values of these two samples with the uncertainty given by the standard error on the median.}
\label{fig:z_mag_VHZ}
\end{figure}

In Fig.~\ref{fig:z_mag_VHZ} we show the 
distribution of apparent magnitude versus photometric redshift of the complete VHZ target sample separated into those with secure redshifts ($z_{\mathrm{q}} \geq3$) and those where we do not measure a secure redshift ($z_{\mathrm{q}} < 3$). (Unlike Fig.~\ref{fig:vhz_zz}-left, here and for all subsequent analysis we use the reference $z_{\mathrm{phot}}$ ``as is'', without forcing it to a high-$z$ solution.) 
The overall photometric redshift distribution of the total VHZ target sample is bimodal with peaks at $z=9-13$ and the corresponding low-z Balmer break solution of $z=0-4$. To ensure robust spectroscopic differentiation between a low-$z$ Balmer break and a high-$z$ Lyman-$\alpha$ break, the VHZ target sample gave higher priority to sources with $m_{1300}\lesssim28$, in order to achieve a PRISM spectrum with $\mathrm{SNR/pix}>2$ just redward of the (expected) Ly-$\alpha$ break. There is little difference in the distribution of photometric redshift between sources where we measure a secure spectroscopic redshift and those where we do not. However, on average the sources where we fail to measure a secure redshift are fainter; most sources with $m_{\mathrm{F277W}}\lesssim27.6$ have secure redshift measurements. 

We do not confirm very high spectroscopic redshifts for any targets with $z_{\mathrm{phot}}>13$. The highest spectroscopic redshift that we measure is $z_{\mathrm{spec}}=12.98$ for UDS-111893, which is discussed further in Appendix~\ref{sec:apdx_uds-z13}.\footnote{CAPERS also observed the now spectroscopically confirmed source MoM-z14 \citep[$z_{\mathrm{spec}}=14.44$; ][]{naidu2025} as COSMOS-102896 but failed to measure a spectroscopic redshift due to the position of the Lyman-$\alpha$ break within a wavelength detector-gap.}

\section{Techniques for selecting galaxies at $\MakeLowercase{z}>9$}
\label{sec:selection_techniques}
In this section we apply a number of commonly used selection techniques to the observed VHZ sample in CAPERS and use the now known spectroscopic redshift to evaluate their effectiveness.  

\subsection{SED fitting templates}
\label{sec:SED_templates}
We first test the ability of different spectral template sets within \textsc{eazy} \citep{brammer2008} to provide accurate photometric redshifts for the selection of galaxies at $z>8.5$ as a wide range of different template sets are used in the literature.  

We first test the ``\textsc{eazy} v1.0'' templates which are the original template set released with \textsc{eazy}. This template set applies the negative matrix factorization (NMF) method from \citet{blanton2007} to reduce the \textsc{pegase} model library \citep{pegase1999} to five templates with the addition of one dusty star-burst template. When fitting using \textsc{eazy}, the best-fitting linear combination of these templates is found. We also test the ``\textsc{eazy} v1.3'' templates, which are an update on the ``\textsc{eazy} v1.0'' templates, including the six original templates updated to include emission lines from \citet{ilbert2009} and with the addition of another three templates. These are a red template from \citet{maraston2005}, a dusty template from \citet{bruzual2003} and the SED of Q2343-BX418 \citep{erb2010}.

We then test the latest templates released with \textsc{eazy} which are the ``tweak\_fsps\_QSF\_v12\_v3'' set of 12 templates generated using Flexible Stellar Population Synthesis (FSPS) models \citep{conroy2010}. These have been ``tweaked'' to best match observed galaxy colors and are the default template set with the latest version of \textsc{eazy}.

We also test the \textsc{pegase 2.0} model library \citep{pegase1999} from \citet{grazian2006} which contains 127 templates that are fitted individually instead of the linear combination typically used with \textsc{eazy}. This template set was used in \citet{donnan2023a,donnan2023b} for the selection of galaxies at $z\geq8.5$.

More recently, \citet{hainline2024b} and \citet{larson2023a} introduced new spectral template sets where older template sets are combined with newer, bluer templates generated to match the UV-colors produced in simulations. \citet{larson2023a} combine the ``tweak\_fsps\_QSF\_v12\_v3'' templates with bluer templates generated using \textsc{bpass} \citep{Eldridge2017} and \textsc{cloudy} \citep{Ferland2017} designed to match the observed UV colors in simulated observations from \citet{yung2022}. \citet{hainline2024b} follow a similar process combining the ``\textsc{eazy} v1.3'' templates with bluer templates generated using FSPS \citep{conroy2010} made to match the observed UV colors in the \textsc{jaguar} simulation \citep{williams2018}. These new template sets have been widely used to select galaxies at $z\geq9$ in \textit{JWST}/NIRCam data \citep[e.g.,][]{finkelstein2022c, finkelstein2024, adams2023, donnan2024, hainline2026}.

\begin{figure*}[ht]
\centering
\includegraphics[width=\linewidth]
{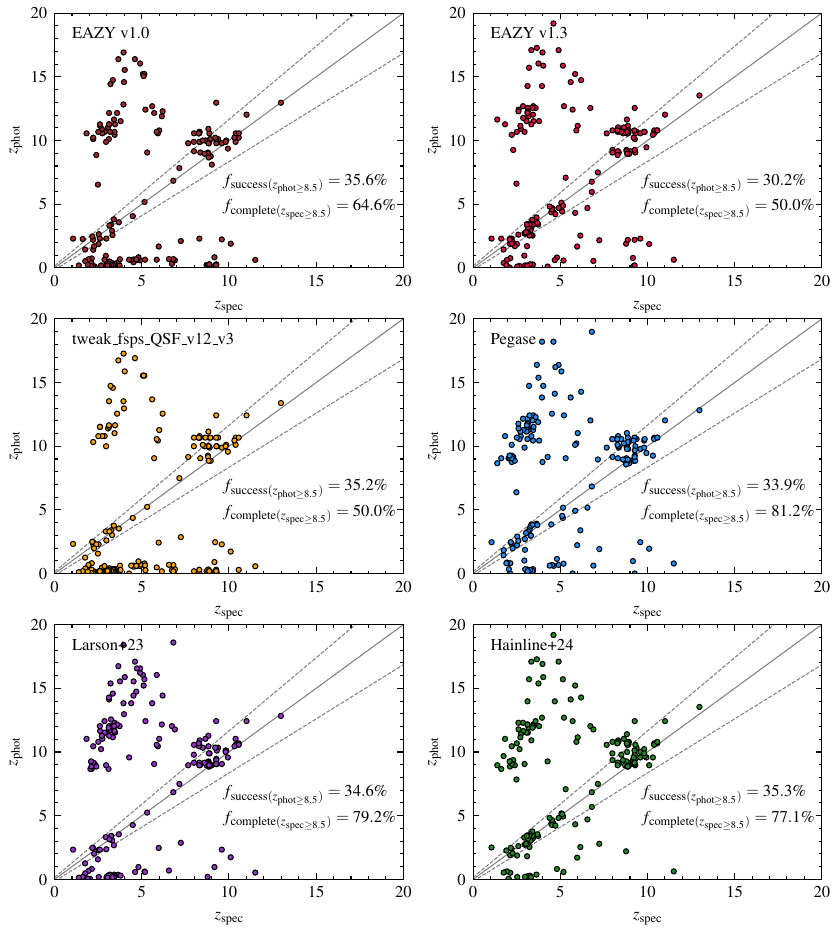}
\caption{A comparison of the photometric redshifts of the 183 CAPERS VHZ targets to the measured spectroscopic redshifts where the photometric redshifts were derived using \textsc{eazy} \citep{brammer2008} with the ``EAZY v1.0'' templates (top left), ``EAZY v1.3'' templates (top right), ``tweak\_fsps\_QSF\_v12\_v3'' templates (center left), ``Pegase'' templates (\citealp{pegase1999}, center right), \citet{larson2023a} templates (bottom left) and the \citet{hainline2024b} templates (bottom right). The inset success rate is defined as the percentage of sources where $|dz|<0.15$, where $dz = (z_{\mathrm{phot}} - z_{\mathrm{spec}})/(1+z_{\mathrm{spec}})$. Sources that fall in the region outlined by the dashed gray lines are those with $|dz|<0.15$.}
\label{fig:templates_zz}
\end{figure*}

We show the success of the different template sets with \textsc{eazy} in Fig.~\ref{fig:templates_zz}. 
The different template sets yield largely similar success rates ranging from $\simeq30\%$ (tweak\_fsps\_QSF\_v12\_v3) to $\simeq35\%$. 
The success rate is determined by two factors:  predicting high photometric redshifts for galaxies with high true redshifts, and by {\em not} falsely predicting high photometric redshifts for foreground, low-redshift galaxies (i.e., minimizing interlopers). This is shown by the completeness rate which ranges from $\simeq50-65\%$ for the redder templates sets but rises significantly to $\simeq80\%$ for the bluer templates sets (Pegase, \citealp{larson2023a}, \citealp{hainline2024}). This demonstrates the importance of including sufficiently blue spectral templates to achieve high-completeness when selecting galaxies at $z\geq8.5$.

Given that the primary photometric redshift measurements used in the target selection for the CAPERS program were determined using the \citet{larson2023a} templates, we adopt these results as our fiducial photometric redshift measurements for the remainder of this paper.

\subsection{Signal-to-noise ratio of detection}
\label{sec:snr}
When selecting galaxy candidates at $z\geq8.5$, the first step is to perform a SNR cut to remove any sources with a weak signal-to-noise ratio. Here we define the SNR of detection as the SNR in the first filter redward of the Lyman-$\alpha$ break given the source's photometric redshift. The most typical cut used is to require a signal $>5\sigma$ \citep[e.g.][]{donnan2023a,donnan2024,leung2023,finkelstein2024,whitler2025} in the filter redward of the break, which varies as a function of redshift. However, some studies use a more conservative $>8\sigma$ requirement instead \citep[e.g.][]{mcleod2024,mcleod2026}.

\begin{figure*}[ht]
\centering
\includegraphics[width=\linewidth]
{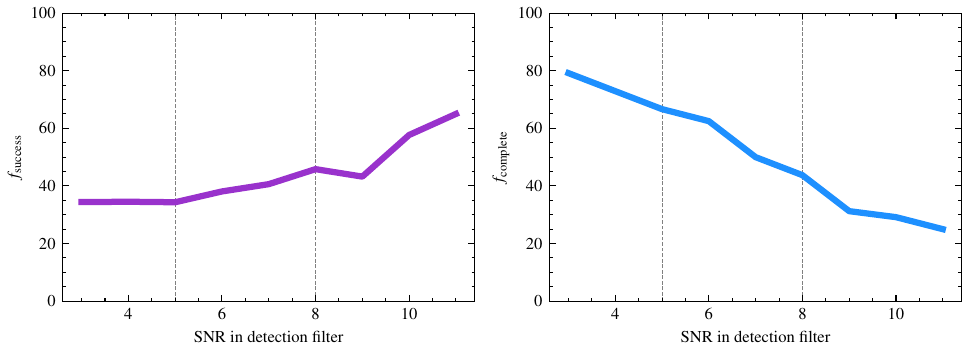}
\caption{\textit{Left}: The percentage of sources at $z_{\mathrm{phot}}\geq8.5$ that are successfully spectroscopically confirmed in the CAPERS program as a function of the minimum signal-to-noise ratio in the detection filter. \textit{Right}: The percentage of sources at $z_{\mathrm{spec}}\geq8.5$ that are successfully selected as a function of the minimum signal-to-noise ratio in the detection filter. The dashed lines in both panels indicate the commonly adopted thresholds of $\mathrm{SNR} \geq 5, 8$.}
\label{fig:snr}
\end{figure*}

To test the impact of different SNR requirements on the ability to successfully select galaxies at $z\geq8.5$, we measure the success rate as a function of SNR in the detection filter. We define the success rate as the fraction of sources at $z_{\mathrm{phot}}\geq8.5$ that have $|dz|<0.15$. The detection filter is dependent on the photometric redshift of the source and is defined as F150W for $8.5<z_{\mathrm{phot}}<9.7$, F200W for $9.7<z_{\mathrm{phot}}<13$ and F277W for $z_{\mathrm{phot}}>13$. We show the results of this in Fig.~\ref{fig:snr}. 

Increasing the SNR requirement increases the success rate of selecting sources at $z\geq8.5$ from $\simeq30\%$ at $\mathrm{SNR}>5$ to $\simeq41\%$ at $\mathrm{SNR}>8$. There is a more significant increase at $\mathrm{SNR}=9-11$ reaching $\simeq60\%$. This is unsurprising given that a higher SNR will select more luminous sources with a stronger Lyman-$\alpha$ break (explored explicitly in Section~\ref{sec:ly_break}). 

However, increasing the SNR can also affect the completeness of the sample where true $z\geq8.5$ sources that have a lower SNR are not selected. We show this in the right-hand panel of Fig.~\ref{fig:snr} where we determine completeness as the fraction of true $z\geq8.5$ sources that are selected by the given SNR requirement. This shows a sharp decline from $\simeq65\%$ at $\mathrm{SNR}=5$ to $\simeq41\%$ at $\mathrm{SNR}=8$. However, this is likely strongly a function of apparent magnitude where faint sources (which are more numerous) fail to be selected by strong SNR requirements.

\subsection{Strength of the Lyman-$\alpha$ break}
\label{sec:ly_break}
\begin{figure*}[ht]
\centering
\includegraphics[width=\linewidth]
{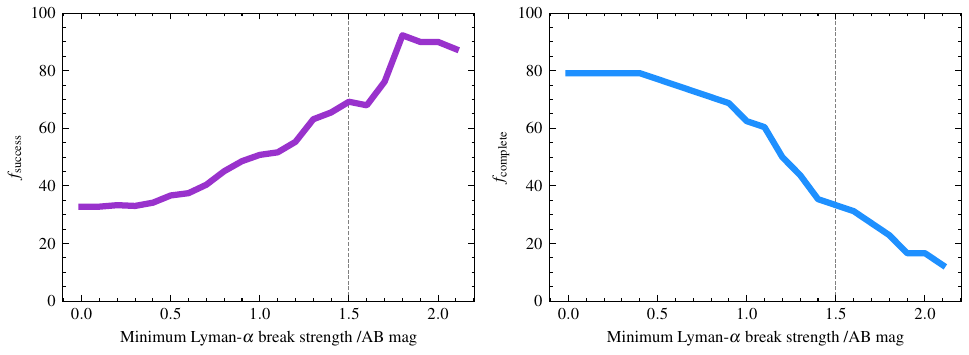}
\caption{\textit{Left}: The percentage of sources at $z_{\mathrm{phot}}\geq8.5$ that are successfully spectroscopically confirmed in the CAPERS program as a function of the Lyman-$\alpha$ break strength. \textit{Right}: The percentage of sources at $z_{\mathrm{spec}}\geq8.5$ that are successfully selected as a function of the Lyman-$\alpha$ break strength. The Lyman-$\alpha$ break strength is defined as the difference in apparent magnitude between the first filter short-ward of $\lambda_{\mathrm{rest}}=1216 \, \mathrm{\AA}$ where the break is at least 90\% through the filter, and the first filter long-ward of $\lambda_{\mathrm{rest}}=1216 \, \mathrm{\AA}$ where the break is not more than 10\% through this filter. The dashed line in both panels indicate a break-strength of $1.5\, \mathrm{mag\, AB}$.}
\label{fig:break_strength}
\end{figure*}

Given that the primary feature that we use to photometrically identify a galaxy at $z\geq8.5$ is the Lyman-$\alpha$ break, the strength of this break can be used to discriminate between secure candidates and low-$z$ contaminants. To measure this we define the break strength as the difference in apparent magnitude between the first filter short-ward of $\lambda_{\mathrm{rest}}=1216 \, \mathrm{\AA}$ where the break is at least 90\% through the filter, and the first filter long-ward of $\lambda_{\mathrm{rest}}=1216 \, \mathrm{\AA}$ where the break is not more than 10\% through this filter, based on the photometric redshift of the source. 
In the case where the flux measurement in a filter has $\mathrm{SNR} < 2$, we adopt a $2\sigma$ upper limit to the flux to set a lower limit on the Lyman-$\alpha$ break strength.
Following the same technique as Section~\ref{sec:snr}, we determine the success rate and completeness rate as a function of the minimum break strength as shown in Fig.~\ref{fig:break_strength}.

This shows a strong improvement in the success rate as a function of increasing break-strength, growing from $\simeq30\%$ at $>0\, \mathrm{mag}$ to $\simeq70\%$ at $>1.5\, \mathrm{mag}$. This is unsurprising given that the maximum strength of a Balmer break (the feature that is confused for a Lyman-$\alpha$ break in low-$z$ contaminants), is strongly unlikely reach strengths beyond $\simeq1.5\, \mathrm{mag}$ in star-forming galaxies \citep{kuruvanthodi2026}. Therefore, if a break in the photometry is measured to be $>1.5\, \mathrm{mag}$, then it is highly probable that it is a Lyman-$\alpha$ break and therefore truly at its high-$z$ photometric redshift. 

However, the completeness rate sharply drops with only $\simeq30\%$ of sources with $z_{\mathrm{spec}}\geq8.5$ having a Lyman-$\alpha$ break $>1.5\, \mathrm{mag}$ in the NIRCam photometry. This is similar to the SNR requirement discussed in Section~\ref{sec:snr} which is strongly a function of source luminosity.  
\textbf{}
\subsection{$\Delta \chi^2$ between low and high redshift solutions}
\label{sec:deltachi2}
The results of SED fitting can not only be used to select those with a preferred redshift of $z>8.5$, but also to evaluate the robustness of this measurement. The goodness-of-fit of an SED template at a given redshift is given by $\chi^2(z)$. For galaxy candidates at $z>8.5$, $\chi^2(z)$ typically has a bi-modal distribution with a trough at the redshift where the break in the photometry is modeled by a Balmer break and another, deeper trough where modeled by the Lyman-$\alpha$ break. Therefore, one can measure the difference in $\chi^2(z)$ between these solutions (where $\Delta \chi^2 = \chi^2_{\mathrm{low\text{-}z}} - \chi^2_{\mathrm{high\text{-}z}}$) and use this to ensure only sources that are secure in their high-$z$ solution are selected. 

\begin{figure*}[ht]
\centering
\includegraphics[width=\linewidth]
{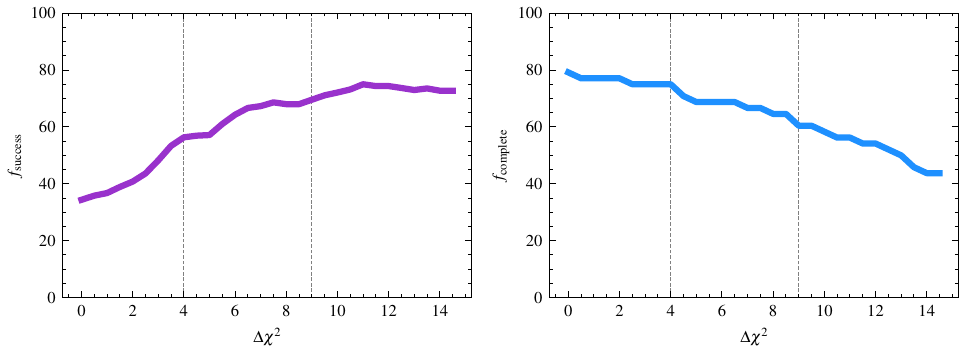}
\caption{\textit{Left}: The percentage of sources at $z_{\mathrm{phot}}\geq8.5$ that are successfully spectroscopically confirmed in the CAPERS program as a function of the minimum value of $\Delta \chi^2$ between the best-fitting low-redshift and high-redshift photometric solution using the \citet{larson2023a} spectral templates with \textsc{eazy} \citep{brammer2008}. \textit{Right}: The percentage of sources at $z_{\mathrm{spec}}\geq8.5$ that are successfully selected as a function of the $\Delta \chi^2$ threshold. The dashed lines in both panels indicate the commonly adopted thresholds of $\Delta \chi^2 \geq 4, 9$.}
\label{fig:chi_2}
\end{figure*}

We measure the impact of different thresholds of $\Delta \chi^2$ in successfully selecting galaxies at $z\geq8.5$. We once again measure the fraction of sources at $z_{\mathrm{phot}}\geq8.5$ with $|dz|<0.15$ as those that are successfully spectroscopically confirmed. We show the results of this in Fig.~\ref{fig:chi_2}.

The success rate of selecting sources at $z\geq8.5$ rises with increasing $\Delta \chi^2$ reaching $\simeq55\%$ at $\Delta \chi^2>4$  and $\simeq70\%$ at $\Delta \chi^2>9$. The completeness decreases with increasing $\Delta \chi^2$ dropping from $\simeq70\%$ at $\Delta \chi^2>4$ to $\simeq60\%$ at $\Delta \chi^2>9$. This is not as strong a decrease as from detection SNR or Lyman-$\alpha$ break strength however suggesting that stringent $\Delta \chi^2$ requirements may be more efficient than stringent SNR or Lyman-$\alpha$ break strength requirements. This is likely because although the $\Delta \chi^2$ is largely driven by the strength of the Lyman-$\alpha$ break, it also is driven by the photometry at $\lambda_{\mathrm{rest}}>1216\, \mathrm{\AA}$ and where a particularly red or blue continuum and the indication of strong rest-frame optical emission lines can help identify low-$z$ contaminants. Therefore, a very strong Lyman-$\alpha$ break (which only the most luminous sources have) is not required to necessarily provide a large $\Delta \chi^2$. We note that this test involves the use of standard NIRCam broad-band filter set (and one medium-band filter; F410M) in CEERS and PRIMER and would likely be more effective with the addition of NIRCam medium-band filters as this would increase the spectral resolution of the SED being modeled with \textsc{eazy}.

\subsection{UV luminosity function redshift prior}
Typical selections of galaxies at $z>9$ assume a flat redshift prior when SED fitting. This assumes that independently of the best photometric redshift inferred from SED fitting, an observed source has an equal probability of being at any redshift. Therefore, the derived $p(z)$ of a given source is simply driven by the observed photometry. However due to the known evolution of the UV LF, a given source has a strongly greater probability of being at $z\sim1-5$ than $z\sim9-12$ with this probability also depending on the implied UV magnitude as a function of redshift \citep[see, e.g.,][]{benitez2000}. \citet{donnan2024} implemented the use of a UV LF prior on the derived $p(z)$ of the selected sources to account for this where the final probability of a source given its flux, $p(z|F)$, is given by 
\begin{equation}
    p(z|F) = \mathcal{L}(F|z) p(z|\mathrm{M_{UV}}),
\end{equation}
where $\mathcal{L}(F|z)$ is the likelihood of the observed fluxes given the redshift,
taken to be
\begin{equation}
    \mathcal{L}(F|z) \propto e^{-\chi^{2}(z)/2},   
\end{equation}
and $p(z|\mathrm{M_{UV}})$ is the prior probability of the redshift based on the implied absolute UV magnitude given by the UV LF prior model. This has the effect of changing the implied photometric redshift of some sources from $z>9$ to $z<6$. 

\begin{figure*}[ht]
\centering
\includegraphics[width=\linewidth]
{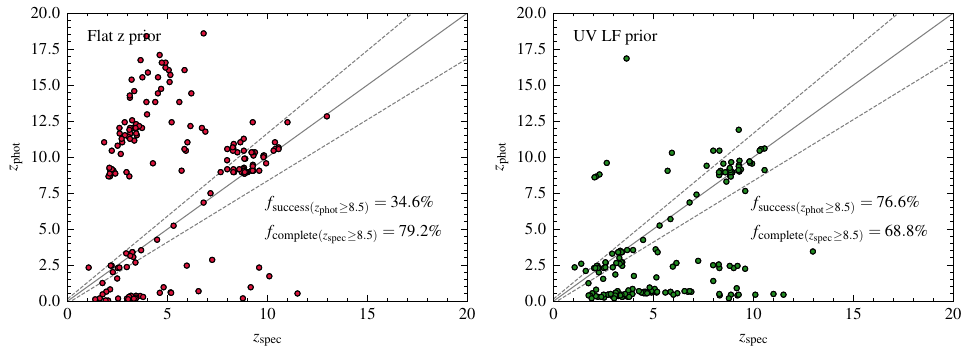}
\caption{A comparison between the spectroscopic and photometric redshifts using \textsc{eazy} \citep{brammer2008} with the \citet{larson2023a} spectral templates with a flat-redshift prior (left) and the UV LF prior from \citet{donnan2024} (right). The inset success rate is defined as the percentage of sources where $|dz|<0.15$, where $dz = (z_{\mathrm{phot}} - z_{\mathrm{spec}})/(1+z_{\mathrm{spec}})$. Sources that fall in the region outlined by the dashed gray lines are those with $|dz|<0.15$.}
\label{fig:UVLF_zz}
\end{figure*}

We use the \citet{donnan2024} UV LF prior on our 183 VHZ targets in CAPERS to test its effectiveness on selecting galaxies at $z>9$. We note that this is different to many of the previous criteria tested (e.g. SNR, $\Delta \chi^2$) which simply further refine a sample after making the photometric redshift measurement. Here the UV LF prior \textit{changes} the initial photometric redshift of each galaxy, due to its impact on their $p(z)$. 
We show the impact of the UV LF prior in Fig.~\ref{fig:UVLF_zz} through a comparison of the photometric and spectroscopic redshifts with simply a flat-redshift prior, to that of the UV LF prior. 

The UV LF prior improves the success rate of selecting galaxies at $z>8$ significantly from $\simeq35\%$ to $\simeq77\%$ while only suffering a moderate drop in completeness ($\simeq79\%$ to $\simeq69\%$). Many of the VHZ candidates that the UV LF removes does so by changing the photometric redshift to $z_{\mathrm{phot}} <2$ which for many sources is now lower than its spectroscopic redshift. However, this is not an issue for the purpose of simply selecting galaxies at $z\geq8.5$ as all that is required is that these low-$z$ interlopers are successfully removed from $z\geq8.5$ samples, which the UV LF prior successfully does.

An obvious concern about the use of a UV LF prior on the selection of $z>9$ galaxies, is that if one is attempting to measure the UV LF, is the final measured UV LF strongly dependent on the assumption of the specific model of the evolving UV LF used for the prior. However, there are two reasons why this is not a significant issue.

Firstly although the use of a UV LF prior does reduce the completeness of the sample, this is corrected by the use of completeness simulations that test the implemented selection process in a given study \citep[e.g.][]{finkelstein2024,donnan2024,whitler2025}. Therefore for example, if a more rapid evolution is assumed in the UV LF then the completeness will be lower but this will be measured by the completeness simulations resulting in a larger correction and a final UV LF measurement that is consistent with other methods.

\begin{figure}[ht!]
\centering
\includegraphics[width=\columnwidth]
{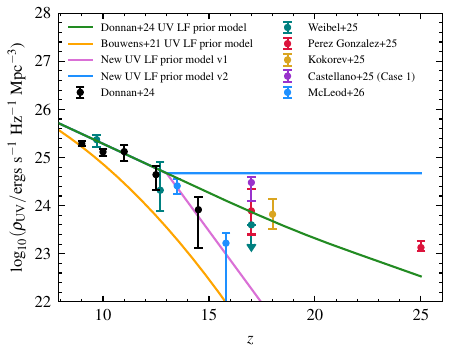}
\caption{The evolution of the cosmic UV luminosity density ($\rho_{\mathrm{UV}}$) shown by observational measurements from \citet{donnan2024, perezgonzalez2025, castellano2025,kokorev2025,weibel2026,mcleod2026} in comparison to the evolving UV LF model from \citet{bouwens2021} (orange), \citet{donnan2024} (green) as well the \citet{donnan2024} model with two alternative versions at $z>13$. The pink line matches the observational measurements showing a more rapid decline at $z>13$ and the blue line shows an extreme example in excess of even the highest measured number densities. }
\label{fig:rho_SFR}
\end{figure} 

Secondly, the exact form for the UV LF at $z>9$ has very little impact on the $p(z)$ given current observational constraints on the UV LF. This can be of particular concern at $z>13$ where there is more uncertainty from observational measurements \citep[e.g.][]{weibel2026,mcleod2026}. To demonstrate that the choice of UV LF evolution at $z>13$ has little impact on the resultant $p(z)$ of $z>13$ candidates, we compare the evolving UV LF model from \citet{donnan2024} with two new versions as well as simply extending the \citet{bouwens2021} model at $z>9$ (although this is disfavored by observational measurements at $z>9$). 

In Fig.~\ref{fig:rho_SFR} we show the UV luminosity density ($\rho_{\mathrm{UV}}$) implied by the UV LF prior model from \citet{donnan2024} in comparison to more recent observational measurements at $z>13$. This model actually is significantly in excess of recent wide-area constraints from \citet{weibel2026,mcleod2026,kreilgaard2026} which imply a more rapid decline in $\rho_{\mathrm{UV}}$. We therefore introduce the first new version of the model to best match this more rapid evolution at $z>13$ and recommend the use of this model for the greatest accuracy in future analysis. This is shown by the pink line in Fig.~\ref{fig:rho_SFR} and is given by the evolving \citet{schechter1976} model from \citet{bouwens2021} at $z<7$, the double-power law model from \citet{donnan2024} at $7<z<13$ and by the following double-power law parameters at $z\geq13$:
\begin{eqnarray*}
    M^{\star} = -19.52\\    
    \phi^{\star} = 10^{(-0.6z+3.62)}\\   
    \alpha = -2.10 \\   
    \beta = -3.43
\end{eqnarray*}
This assumes a pure $\phi^{\star}$ evolution in the UV LF with $M^{\star}, \alpha$ and $\beta$ fixed at $z\geq13$.

The second new version (shown by the blue line in Fig.~\ref{fig:rho_SFR}) assumes an arbitrarily high $\rho_{\mathrm{UV}}$ at $z>13$ which is in significant excess of the observational measurements. This is an extreme example to test if even such an extreme model has a significant impact on the resultant $p(z)$ for a $z>13$ candidate.

We find negligible difference in the completeness of our $z>9$ sample between the \citet{donnan2024} model and the new v1 and v2 models of the UV LF prior, finding only one $z\simeq17$ candidate (COSMOS-39810, $z_{\mathrm{spec}}=3.66$) for which the \citet{donnan2024} UV LF prior failed to drive to low-$z$. The new $z>13$ rapid-decline model now increases the $p(z<6)$ sufficiently that the source is no longer preferred at $z>13$. The \citet{bouwens2021} model has a small impact with 4 additional sources moving to their low-$z$ solutions. These are sources for which the low-redshift and high-redshift peaks in the $p(z)$ are similar and therefore the difference from the UV LF prior model does impact them. However as shown in Fig.~\ref{fig:rho_SFR}, \textit{JWST} observations of the UV LF at $z>9$ disfavor this model.

\begin{figure*}[ht]
\centering
\includegraphics[width=\linewidth]
{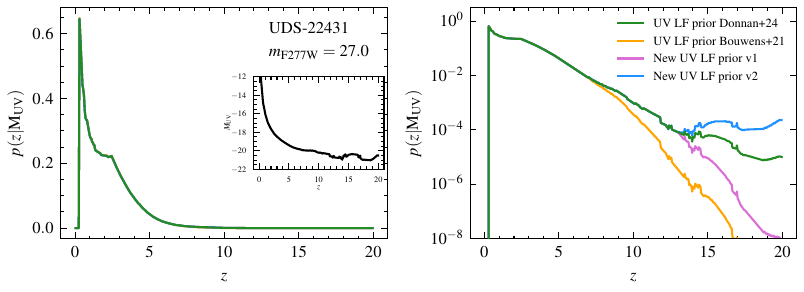}
\caption{\textit{Left}: The redshift evolution of the $p(z|\mathrm{M_{UV}})$ for UDS-22431 given the $M_{\mathrm{UV}}(z)$ shown by the inset panel (derived using the SED fitting code from \citet{mclure2011}). This is shown using the UV LF model from \citet{donnan2024} (green), \citet{bouwens2021} (orange) and the alternative \citet{donnan2024} model at $z>13$ where there is a rapid decline (pink) and an extreme lack of evolution (blue). \textit{Right}: The same as left but with a log-scale revealing the small differences between the models which are at a level of $<1\%$.}
\label{fig:LF_pz_example}
\end{figure*}

The explanation for the lack of impact of the difference in the resultant $p(z)$ from these models, is that the probability is significantly higher at $z\sim1-5$ than at $z>9$ due to the orders-of-magnitude greater number density of galaxies at  $z\sim1-5$ than at $z>9$ and the fainter implied UV magnitude of a source at low-$z$. Therefore, the probability of a source given the evolving UV LF is already so low at $z>9$ in comparison to $z\sim1-5$, that differences between models at $z>9$ are extremely insignificant. We demonstrate this for an example source (UDS-22431) which has an apparent magnitude typical for luminous $z>9$ candidates ($m_{\mathrm{F277W}}=27.0$), in Fig.~\ref{fig:LF_pz_example}. The difference between these models is not apparent by eye unless plotted on a log-scale as the difference in $p(z)$ is $<1\%$ at $z>9$ demonstrating the extremely low impact of the choice between these models in the resultant $p(z)$ when selecting galaxies at $z>9$. 

\subsection{Recommendations for effective selection of galaxies at $z\geq9$}

The ideal method of selecting $z\geq8.5$ galaxies would yield a 100\% success rate with 100\% completeness by selecting all the sources truly at $z\geq8.5$ with no contamination. In reality all selection methods fail to truly remove all low-$z$ contaminants and simultaneously falsely categorize some true $z\geq8.5$ as low-$z$ contaminants as we have shown in Section~\ref{sec:selection_techniques}.

In particular a selection simply based on the strength of the Lyman-$\alpha$ break alone, although highly correlated with redshift success, has a low completeness with only $\simeq30\%$ of the confirmed $z\geq8.5$ sources in CAPERS being selected with a break-strength of $>1.5\ \mathrm{mag}$. Using the results of SED fitting with a $\Delta \chi^2$ cut is able to more efficiently select sources at $z\geq8.5$. 

The most common selection method involves a combination of a SNR and a $\Delta \chi^2$ requirement. Here we test the combination of these criteria to determine raising which leads to a more successful final outcome. The typical minimum criteria is $\mathrm{SNR}>5$ and $\Delta \chi^2\geq4$. If we raise the SNR requirement to $\mathrm{SNR}>8$ keeping a requirement of $\Delta \chi^2>4$, we obtain a success rate of $\simeq65\%$ with a completeness of $\simeq37\%$. Conversely, if we keep a requirement of $\mathrm{SNR}>5$ but use a more strict requirement of $\Delta \chi^2>9$, we reach a success rate of $\simeq69\%$ and a completeness of $\simeq51\%$. This demonstrates that if one implements a selection method based on these two criteria, it is more effective to use a strict $\Delta \chi^2$ requirement instead of a strict $\mathrm{SNR}$ requirement given this yields a higher success rate and improved completeness.

The use of a UV LF prior over the typically assumed flat redshift prior is even more efficient at removing low-$z$ contaminants. We have shown that with a UV LF prior the success rate rises to $\sim77\%$ compared to $\sim35\%$ with a flat redshift prior. As just demonstrated, this high success rate is even greater than that achieved by combining an SNR requirement of $>5\sigma$ and a $\Delta \chi^2>9$ ($f_{\mathrm{success}}\simeq69\%$). Furthermore, the completeness is significantly higher than an SNR requirement of $>5\sigma$ and a $\Delta \chi^2>9$ ($\simeq69\%$ compared to $\simeq51\%$). Therefore simply the use of a UV LF prior in deriving photometric redshifts is significantly the most effective selection technique tested in this work. We have also demonstrated that given current observational constraints of the UV LF at $z=9-14$, the derived photometric redshifts of high-$z$ sources is invariant to precise model of the evolving UV LF. 

\section{Nature of low-redshift interlopers}
\label{sec:z_failure}
\begin{figure}[ht]
\centering
\includegraphics[width=\columnwidth]
{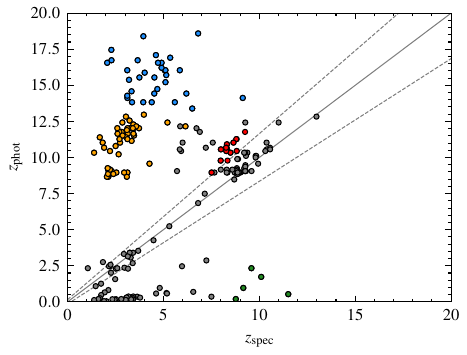}
\caption{A comparison of the photometric redshifts of the 183 CAPERS VHZ targets to the measured spectroscopic redshifts where the photometric redshifts were derived using \textsc{eazy} \citep{brammer2008} with the \citet{larson2023a} templates. A redshift success is defined as $|dz|<0.15$, where $dz = (z_{\mathrm{phot}} - z_{\mathrm{spec}})/(1+z_{\mathrm{spec}})$. Sources that fall in the region outlined by the dashed gray lines are those with $|dz|<0.15$. We show three sub-samples of sources with redshift failures which are ``near-misses'' (red), low-$z$ galaxies with $8.5\leq z_{\mathrm{phot}}\leq 13$ (orange) and low-$z$ galaxies with $z_{\mathrm{phot}}>13$ (blue). }
\label{fig:z_failures}
\end{figure}

Given there are a number of VHZ candidates that fail to be confirmed as $z\geq8.5$ galaxies, we investigate the specific properties of these redshift failures in this section. We define three different categories of redshift failures shown in Fig.~\ref{fig:z_failures} in which we investigate their observational and physical properties.

We measure the implied rest-frame UV continuum slope of each VHZ candidate where $\beta$ is given by $f_{\lambda}\propto\lambda^{\beta}$. We fit to the photometric data-points with this power-law following the method of \citet{cullen2023} where redshift is fixed to the photometric redshift. This enables a test of whether the successful VHZ candidates differ in their implied UV slopes (which are incorrect in the case of low-redshift contaminants) to those candidates that fail.  

\begin{figure}[ht]
\centering
\includegraphics[width=\columnwidth]
{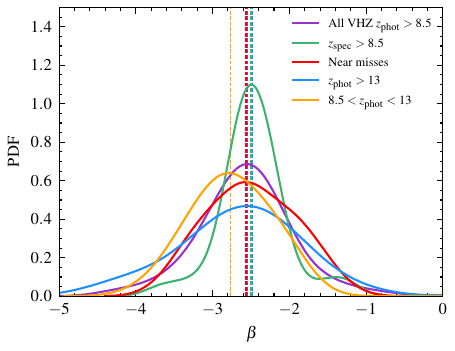}
\caption{The probability density function (PDF) for the UV-continuum slope ($\beta$) measured from NIRCam photometry using the method from \citet{cullen2023} for every VHZ candidate at $z_{\mathrm{phot}}\geq8.5$ (purple). We also show the PDF for only those successfully spectroscopically confirmed to be at $z\geq8.5$ (green). We follow the classification in Fig.~\ref{fig:z_failures} for those that fail to be spectroscopically confirmed to be $z\geq8.5$ (red, blue and orange). In all cases the UV-slope fit uses a fixed redshift given by the photometric redshift.}
\label{fig:beta_kde}
\end{figure}

In Fig.~\ref{fig:beta_kde} we show the probability density functions for $\beta$ firstly for the full sample of VHZ candidates that have $z_{\mathrm{phot}}\geq8.5$ (purple), then for those sources that are successfully spectroscopically confirmed (green) followed by the three categories of redshift failures defined in Fig.~\ref{fig:z_failures}.
The total VHZ candidate sample spans a wide range of UV slopes from $\beta=-5$ to $\beta=-1$ with a median of $\beta\simeq-2.5$. Those sources successfully spectroscopically confirmed have a similar median but occupy a significantly narrower range of UV slopes with most of the probability in between $\beta=-3$ and $\beta=-1$. The ``near-misses'' and $z_{\mathrm{phot}}\geq13$ candidates have similar median UV slopes in contrast to the $8.5\leq z_{\mathrm{phot}}\leq 13$ candidates which have a bluer median UV slope of $\beta\simeq-2.8$ with a significant tail to very blue UV slopes. We discuss this in further detail in Section~\ref{sec:8.5_contaminants}. Given the narrower range of $\beta$ for the redshift successes, this suggests that extremely blue UV slopes ($\beta \ll-3$) in particular could be a signature of low-redshift contamination.

\subsection{Near-misses}
\label{sec:near_misses}
Firstly, we define ``near-misses'' which are sources for which are confirmed high-$z$ but have lower spectroscopic redshifts such that it does not meet the $|dz|<0.15$ definition of redshift success. We define these sources as those with $z_{\mathrm{phot}}\geq8.5$ and $0.15<|dz|<0.3$ with these 14 sources shown in red in Fig.~\ref{fig:z_failures}. These sources occupy a specific redshift range and all have elevated F410M and/or F444W fluxes relative to F277W and F356W. When SED fitting with \textsc{eazy} to obtain a photometric redshift, this rise in flux in these filters is often interpreted as a Balmer break rather than strong $\mathrm{H} \beta$ and \oiiiboth\, emission. This has the effect of increasing the photometric redshift by $\Delta z\simeq2$ compared to the true redshift. Given that $\mathrm{H} \beta$ and \oiiiboth\, are redshifted beyond $\lambda_{\mathrm{obs}}\simeq5\, \mu\mathrm{m}$ at $z>9$, this effect does not occur for sources at $z_{\mathrm{spec}}\gtrsim9$.

This effect is likely a product of the method by which \textsc{eazy} obtains a best-fitting SED model. \textsc{eazy} combines multiple templates in a linear way to simply best-match the photometry rather than fitting a model parameterized by physical properties. This can lead to large Balmer breaks in the best fitting \textsc{eazy} model which are physically disfavored. 

\subsection{Low-$z$ contaminants at $8.5\leq z_{\mathrm{phot}}\leq13$}
\label{sec:8.5_contaminants}
Next we select sources of redshift failure with high photometric redshift $8.5\leq z_{\mathrm{phot}}\leq13$ but low spectroscopic redshift $z_{\mathrm{spec}}\lesssim6$. We select a sample of 49 sources which have $8.5\leq z_{\mathrm{phot}}\leq13$ and $|dz|>0.8$ which are shown by the orange points in Fig.~\ref{fig:z_failures}. These are typical low-$z$ contaminants where their Balmer break (as revealed by spectroscopy) were confused for a Lyman-$\alpha$ break at high-$z$. As noted in Fig.~\ref{fig:beta_kde}, the average UV-continuum slope of these sources is bluer than the total VHZ candidate sample and crucially bluer than the confirmed sources at $z\geq8.5$. This is likely due to the NIRCam photometry being dominated by strong rest-frame optical emission lines at their true redshift of $2 \lesssim z_{\mathrm{spec}} \lesssim 4$ which imitate an extremely blue UV continuum at $z\geq8.5$. For example, the $z_{\mathrm{phot}}\simeq12$ candidate EGS-55434 which at its photometric redshift has an implied UV slope of $\beta = -3.4\pm0.5$, has a secure spectroscopic redshift of $z_{\mathrm{spec}}=3.434$. At this spectroscopic redshift, the strong \oiiiboth\, lines dominate the flux in the F200W filter with H$\alpha$ dominating the flux in F356W. Since \oiiiboth\, is significantly stronger than H$\alpha$, this creates the appearance of a blue color between these filters. 

However, although these candidates have bluer UV slopes than those with successful $z\geq8.5$ confirmations, the UV slope of a VHZ candidate is not a ``smoking gun'' signature of low-redshift contamination. This is because UV slopes can be quite uncertain given the low number of data-points and broad wavelength range of the wide NIRCam filters (e.g. the uncertainty on EGS-55434 of $\Delta \beta =0.5$), and that their distribution overlaps significantly with those that are successfully confirmed to be $z\geq8.5$. Therefore, it is only most useful in cases where the implied UV slopes are precise enough and blue enough ($\beta \ll -3$), such that they are physically incompatible with a $z\geq8.5$ solution.

\subsection{Low-$z$ contaminants at $z_{\mathrm{phot}}>13$}
Finally, we investigate those candidates at $z_{\mathrm{phot}}>13$ of which none were successfully spectroscopically confirmed. We define this sub-sample as those with $z_{\mathrm{phot}}>13$ and $|dz|>0.15$ of which there are 33 sources shown by the blue points in Fig.~\ref{fig:z_failures}. 

As shown in Fig.~\ref{fig:vhz_zz}, the vast majority of these sources have a VHZ count $<3$. As described in Section~\ref{sec:vhz_selection}, high VHZ count strongly correlates with redshift confirmation success. The lack of successful redshift confirmation is then perhaps unsurprising given the less secure nature of these candidates. 

Out of the $z_{\mathrm{phot}}>13$ candidates observed with CAPERS, five were identified and their spectroscopic redshifts reported in \citet{castellano2025} ranging from $z\simeq 3-6.5$. These sources can be categorized as either dusty emission-line galaxies or low-mass passive galaxies.

\section{Confirmed sources at $\MakeLowercase{z}\geq9$ in CAPERS}
\label{sec:conf_VHZ}
\subsection{Observational properties}
Out of the 183 sources classified as VHZ candidates in CAPERS, we now isolate the successful confirmations of galaxies at $z>8.5$. Given that the focus of this paper is the evaluation of selection methodology, we refrain from a detailed analysis of these sources here. Instead we provide their basic observational properties and leave a more detailed analysis for future work.

The observational properties of the 48 sources we successfully confirm are listed in Table~\ref{tab:sample}-\ref{tab:sample2}. Here we note the spectroscopic redshift and the respective quality flag $z_{\mathrm{q}}$ of each source. A number of these sources have been identified in previous studies which we note in the right-hand column. 
\begin{figure*}[ht]
\centering
\includegraphics[width=\linewidth]
{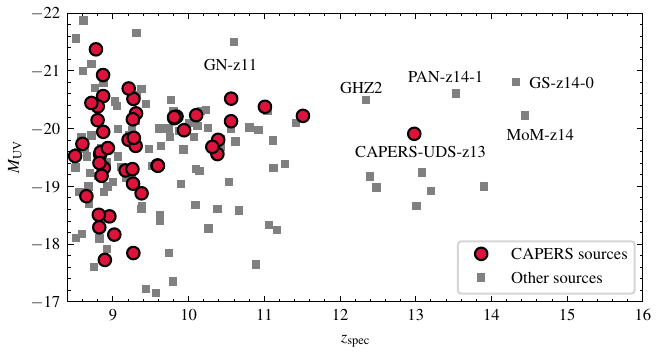}
\caption{The spectroscopic redshifts and absolute UV magnitudes of the confirmed CAPERS sources at $z_{\mathrm{spec}}\geq8.5$ (red) in comparison to all the other sources in the literature. Labeled are the notable luminous sources of CAPERS-UDS-z13 (UDS-111893), GN-z11 \citep{bunker2023}, GHZ2 \citep{castellano2024}, GS-z14-0 \citep{carniani2024}, MoM-z14 \citep{naidu2025} and PAN-z14-1 \citep{donnan2026}.}
\label{fig:Muv_z} 
\end{figure*}

\begin{deluxetable*}{lcccccccc}
\caption{Observational properties of the 48 spectroscopically confirmed galaxies at $z\geq8.5$ in CAPERS.}
\label{tab:sample}
\tablehead{ID & RA & DEC & $z_{\mathrm{phot}}$ & $z_{\mathrm{spec}}$ & $z_{\mathrm{q}}$ & $M_{\mathrm{UV}}$ & $\beta$ & Reference \\ 
& /deg & /deg & & & & /AB mag & &   }

\startdata
UDS-74581 & 34.515057 & -5.261461 & 10.36 & $8.5003^{+0.0012}_{-0.0012}$ & 4 & $-$19.53 & $-2.24^{+0.69}_{-0.55}$ & - \\
UDS-132353 & 34.214517 & -5.112473 & 9.13 & $8.5973^{+0.0088}_{-0.0088}$ & 4 & $-$19.74 & $-2.42^{+0.20}_{-0.21}$ & - \\
COSMOS-116486 & 150.176036 & 2.301991 & 10.99 & $8.650^{+0.014}_{-0.014}$ & 3 & $-$18.83 & $-2.25^{+0.34}_{-0.32}$ & - \\
EGS-11964 & 214.968697 & 52.929651 & 8.92 & $8.7158^{+0.0035}_{-0.0035}$ & 4 & $-$20.44 & $-2.49^{+0.16}_{-0.15}$ & - \\
EGS-28945 & 215.008658 & 52.868294 & 2.14 & $8.7760^{+0.0035}_{-0.0035}$ & 4 & $-$21.38 & $-2.57^{+0.20}_{-0.22}$ & - \\
EGS-4347 & 214.994395 & 52.989380 & 9.07 & $8.7977^{+0.0033}_{-0.0033}$ & 4 & $-$20.15 & $-2.64^{+0.13}_{-0.12}$ & H24 \\
EGS-3044 & 214.985106 & 52.990468 & 9.10 & $8.8004^{+0.0030}_{-0.0030}$ & 4 & $-$20.38 & $-2.13^{+0.25}_{-0.23}$ & - \\
EGS-87149 & 214.818301 & 52.834279 & 9.07 & $8.8138^{+0.0016}_{-0.0016}$ & 4 & $-$18.51 & $-2.19^{+0.21}_{-0.22}$ & - \\
EGS-64034 & 214.871721 & 52.833797 & 11.26 & $8.82^{+0.08}_{-0.07}$ & 4 & $-$18.29 & $-1.92^{+0.77}_{-0.65}$ & - \\
UDS-89601 & 34.418919 & -5.227818 & 10.45 & $8.8225^{+0.0033}_{-0.0033}$ & 4 & $-$19.41 & $-2.70^{+0.68}_{-0.62}$ & N26 \\
EGS-85546 & 214.938798 & 52.922734 & 9.28 & $8.8311^{+0.0038}_{-0.0038}$ & 4 & $-$19.55 & $-2.51^{+0.28}_{-0.26}$ & - \\
EGS-34935 & 214.996395 & 52.854155 & 8.71 & $8.838^{+0.013}_{-0.013}$ & 4 & $-$19.59 & $-2.71^{+0.20}_{-0.17}$ & - \\
COSMOS-168193 & 150.160277 & 2.464053 & 9.85 & $8.8504^{+0.0013}_{-0.0013}$ & 4 & $-$19.18 & $-2.22^{+0.17}_{-0.15}$ & - \\
EGS-6691 & 215.011704 & 52.988304 & 8.98 & $8.8710^{+0.0020}_{-0.0020}$ & 4 & $-$20.56 & $-2.61^{+0.08}_{-0.08}$ & H24 \\
COSMOS-17314 & 150.126367 & 2.383787 & 8.95 & $8.8699^{+0.0038}_{-0.0038}$ & 4 & $-$20.93 & $-2.45^{+0.07}_{-0.07}$ & - \\
COSMOS-109633 & 150.065629 & 2.287479 & 8.80 & $8.8721^{+0.0018}_{-0.0018}$ & 4 & $-$19.95 & $-2.84^{+0.13}_{-0.13}$ & - \\
EGS-61226 & 214.901242 & 52.846996 & 8.92 & $8.8794^{+0.0028}_{-0.0028}$ & 4 & $-$19.32 & $-2.23^{+0.42}_{-0.37}$ & H24 \\
EGS-36064 & 214.975841 & 52.841949 & 9.91 & $8.8936^{+0.0020}_{-0.0020}$ & 4 & $-$17.73 & $-1.05^{+0.29}_{-0.26}$ & - \\
COSMOS-153965 & 150.159776 & 2.379294 & 8.95 & $8.9333^{+0.0020}_{-0.0020}$ & 4 & $-$19.66 & $-2.38^{+0.18}_{-0.16}$ & - \\
EGS-55806 & 214.958341 & 52.872524 & 8.95 & $8.954^{+0.014}_{-0.014}$ & 4 & $-$18.48 & $-2.49^{+0.30}_{-0.27}$ & - \\
EGS-94554 & 214.878012 & 52.856076 & 8.98 & $9.0183^{+0.0014}_{-0.0014}$ & 4 & $-$18.24 & $-2.72^{+0.35}_{-0.33}$ & MT26 \\
COSMOS-110865 & 150.083362 & 2.290118 & 0.97 & $9.167^{+0.030}_{-0.030}$ & 3 & $-$19.35 & $-1.08^{+0.89}_{-0.60}$ & - \\
EGS-48895 & 214.989020 & 52.879277 & 9.04 & $9.2072^{+0.0020}_{-0.0020}$ & 4 & $-$19.36 & $-2.46^{+0.34}_{-0.32}$ & - \\
UDS-28597 & 34.465142 & $-$5.217735 & 10.30 & $9.2083^{+0.0023}_{-0.0023}$ & 4 & $-$20.71 & $-2.12^{+0.24}_{-0.24}$ & MT26 \\
UDS-146485 & 34.249791 & $-$5.142105 & 9.07 & $9.264^{+0.016}_{-0.016}$ & 3 & $-$20.05 & $-2.50^{+0.13}_{-0.14}$ & MT26 \\
UDS-142042 & 34.249398 & $-$5.130833 & 9.10 & $9.2581^{+0.0019}_{-0.0019}$ & 3 & $-$19.34 & $-2.49^{+0.46}_{-0.45}$ & MT26 \\
UDS-22431 & 34.460257 & $-$5.185003 & 9.31 & $9.2710^{+0.0037}_{-0.0037}$ & 4 & $-$20.52 & $-2.33^{+0.08}_{-0.08}$ & P26, MT26 \\
COSMOS-153330 & 150.174418 & 2.377952 & 9.07 & $9.266^{+0.013}_{-0.013}$ & 4 & $-$19.12 & $-2.21^{+0.30}_{-0.28}$ & - \\
UDS-127376 & 34.436059 & $-$5.101416 & 9.49 & $9.276^{+0.015}_{-0.015}$ & 3 & $-$19.80 & $-2.34^{+0.10}_{-0.09}$ & MT26 \\
COSMOS-119334 & 150.136253 & 2.308030 & 11.92 & $9.288^{+0.003}_{-0.003}$ & 3 & $-$17.79 & $-0.50^{+0.42}_{-0.37}$ & AT25 \\
UDS-60476 & 34.520975 & $-$5.292676 & 10.00 & $9.3054^{+0.0021}_{-0.0021}$ & 4 & $-$20.25 & $-1.93^{+0.11}_{-0.11}$ & - \\
COSMOS-118821 & 150.090288 & 2.306891 & 9.01 & $9.300^{+0.017}_{-0.017}$ & 4 & $-$19.86 & $-2.76^{+0.17}_{-0.16}$ & MT26 \\
EGS-87132 & 215.044005 & 52.994301 & 10.33 & $9.37878^{+0.00090}_{-0.00090}$ & 4 & $-$18.85 & $-2.25^{+0.24}_{-0.24}$ & P26, MT26 \\
COSMOS-38588 & 150.143960 & 2.277489 & 2.32 & $9.59^{+0.11}_{-0.11}$ & 9 & $-$19.30 & $-1.24^{+0.31}_{-0.29}$ & - \\
UDS-96334 & 34.293454 & $-$5.212675 & 9.04 & $9.59470^{+0.00039}_{-0.00039}$ & 4 & $-$19.51 & $-2.57^{+0.52}_{-0.48}$ & - \\
UDS-70629 & 34.466897 & $-$5.269839 & 10.12 & $9.84^{+0.03}_{-0.03}$ & 3 & $-$20.20 & $-2.57^{+0.14}_{-0.13}$ & - \\
COSMOS-300001 & 150.1165406 & 2.1971091 & 9.73 & $9.8083^{+0.0072}_{-0.0072}$ & 4 & $-$20.35 & $-2.82^{+0.29}_{-0.31}$ & D25 \\
COSMOS-52185 & 150.116567 & 2.197055 & 9.73 & $9.8097^{+0.0034}_{-0.0034}$ & 4 & $-$20.33 & $-2.55^{+0.18}_{-0.18}$ & D25, MT26 \\
EGS-25297 & 214.817106 & 52.748342 & 9.58 & $9.939^{+0.031}_{-0.031}$ & 4 & $-$19.93 & $-1.66^{+0.09}_{-0.09}$ & D25, P26, MT26 \\
\enddata
\tablecomments{References: H24 \citep{harikane2024}, D25 \citep{donnan2025b}, N26 \citep{napolitano2026}, W26 \citep{weibel2026}, P26 \citep{pollock2025}, AT25 \citep{taylor2025}, K25 \citep{kokorev2025}, MT26 \citep{tang2026}, RB26 \citep{robertsborsani2026}, H26 \citep{heintz2026}.}
\end{deluxetable*}

\begin{deluxetable*}{lcccccccc}
\caption{Table~\ref{tab:sample} continued.}
\label{tab:sample2}
\tablehead{ID & RA & DEC & $z_{\mathrm{phot}}$ & $z_{\mathrm{spec}}$ & $z_{\mathrm{q}}$ & $M_{\mathrm{UV}}$ & $\beta$ & Reference \\ 
& /deg & /deg & & & & /AB mag & &   }

\startdata
COSMOS-107634 & 150.161687 & 2.283384 & 10.72 & $10.10^{+0.11}_{-0.12}$ & 9 & $-$20.34 & $-1.78^{+0.73}_{-0.50}$ & - \\
UDS-82699 & 34.413555 & $-$5.244245 & 10.45 & $10.313^{+0.021}_{-0.021}$ & 3 & $-$19.64 & $-2.22^{+0.21}_{-0.21}$ & W26, H26 \\
COSMOS-109917 & 150.142909 & 2.288019 & 11.05 & $10.38^{+0.06}_{-0.06}$ & 3 & $-$19.65 & $-2.82^{+0.30}_{-0.26}$ & MT26, RB26, H26 \\
EGS-54833 & 214.885966 & 52.819062 & 10.54 & $10.39^{+0.06}_{-0.06}$ & 4 & $-$19.57 & $-2.60^{+0.32}_{-0.31}$ & - \\
EGS-22637 & 214.932075 & 52.841871 & 10.69 & $10.56^{+0.05}_{-0.05}$ & 9 & $-$20.09 & $-2.47^{+0.33}_{-0.31}$ & RB26, H26 \\
UDS-136645 & 34.456023 & $-$5.121952 & 9.07 & $10.562^{+0.034}_{-0.034}$ & 3 & $-$20.56 & $-2.61^{+0.61}_{-0.54}$ & K25, MT26, RB26, H26 \\
UDS-126973 & 34.264439 & $-$5.096232 & 12.22 & $11.008^{+0.017}_{-0.017}$ & 3 & $-$20.44 & $-1.78^{+0.16}_{-0.15}$ & K25, MT26, RB26, H26 \\
UDS-88619 & 34.319134 & $-$5.230137 & 0.55 & $11.51^{+0.04}_{-0.04}$ & 3 & $-$20.28 & $-1.97^{+0.44}_{-0.37}$ & H26 \\
UDS-111893 & 34.292010 & $-$5.174140 & 12.76 & $12.98^{+0.18}_{-0.16}$ & 9 & $-$20.03 & $-2.42^{+0.34}_{-0.35}$ & - \\
\enddata
\tablecomments{References: H24 \citep{harikane2024}, D25 \citep{donnan2025b}, N26 \citep{napolitano2026}, W26 \citep{weibel2026}, P26 \citep{pollock2025}, AT25 \citep{taylor2025}, K25 \citep{kokorev2025}, MT26 \citep{tang2026}, RB26 \citep{robertsborsani2026}, H26 \citep{heintz2026}.}
\end{deluxetable*}

We measure the rest-frame UV absolute magnitude of the galaxies at $z_{\mathrm{spec}}\geq8.5$ in CAPERS by first modeling the photometry of each source with \textsc{bagpipes} \citep{carnall2018}. We do this by fitting the \textit{JWST}/NIRCam and \textit{HST}/ACS photometry fixing the redshift to the spectroscopic measurement. We then integrate the flux given by the best-fitting model over a top-hat function centered on $\lambda_{\mathrm{rest}}=1500\, \mathrm{\AA}$ with a width of $100\, \mathrm{\AA}$. 
The UV absolute magnitudes of the 48 confirmed sources at $z>8.5$ are listed in Tables~\ref{tab:sample}-\ref{tab:sample2} and are shown in Fig.~\ref{fig:Muv_z} in comparison to the remaining measurements of spectroscopically confirmed sources from the literature. 

\begin{figure}[ht]
\centering
\includegraphics[width=\columnwidth]
{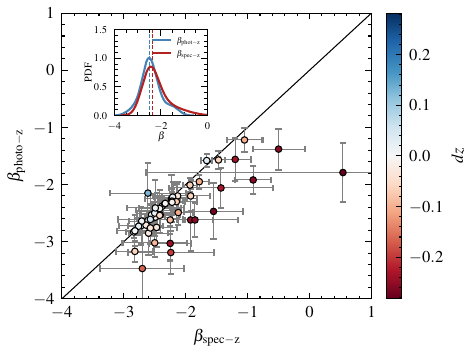}
\caption{The measured UV-continuum slopes of the 48 spectroscopically confirmed sources at $z\geq8.5$ in CAPERS from fitting NIRCam photometry where the redshift is fixed to the photometric redshift in comparison to when fixed to the spectroscopic redshift. The color of the data-points shows the redshift accuracy where $dz=(z_{\mathrm{phot}} - z_{\mathrm{spec}})/(1+z_{\mathrm{spec}})$. \textit{Inset}: The probability density function (PDF) of each axis where the dashed line shows the median of the distribution.}
\label{fig:beta_comparison}
\end{figure}

We once again measure the rest-frame UV continuum slopes, $\beta$, by fitting to the photometric data-points of the confirmed VHZ sample following the method of \citet{cullen2023} where redshift is now fixed to the spectroscopic redshift. We list the results in Tables~\ref{tab:sample}-\ref{tab:sample2} for each galaxy.  We then compare their UV slope measurements to the previous measurements where the redshift was fixed to the photometric value as shown in Fig.~\ref{fig:beta_comparison}.

The majority of the sample has consistent $\beta$ values when the redshift is either the photometric or spectroscopic measurement. However, there are a number of sources which have redder UV slopes when using the spectroscopic redshift. As shown by the color of the data-points (given by $dz$), the comparison in $\beta$ is strongly correlated by how consistent the redshift measurements are with redder UV slopes in the cases where the photometric redshift was overestimated compared to the spectroscopic redshift. 

Similar to other works \citep[e.g.][]{serjeant2023,arrabalharo2023b, hainline2024,heintz2026}, we find a systematic overestimation in photometric redshifts of $\langle \Delta z\rangle=0.37$ among the spectroscopically confirmed sources at $z_{\mathrm{spec}}\geq8.5$ with $|dz|<0.15$.
This average photometric redshift overestimation drives a systematic bias towards bluer UV slopes when measured using photometric redshifts, than in reality becomes apparent with the use of the spectroscopic redshifts. In our sample this changes the median UV-slope from $\beta\simeq-2.5$ to $\beta\simeq-2.3$.

\subsection{Notable sources}
The CAPERS confirmed VHZ sample contains some notable sources that have been discussed in further detail in other studies. Firstly, UDS-136645 and UDS-126973 were first presented in \citet{kokorev2025} (labeled CAPERS\_UDS\_z10 and CAPERS\_UDS\_z11 respectively). COSMOS-119334 is currently the highest known redshift broad-line active galactic nuclei (AGN) and little red dot (LRD) at $z_{\mathrm{spec}}=9.288$ \citep[CAPERS-LRD-z9;][]{taylor2025}. The SED of this source is best-modeled by a dense gas enshrouded black hole as this gives rise to the extremely strong Balmer break similar to other extreme LRDs \citep[e.g.][]{degraaff2025,naidu2025b}. This is the only source in the CAPERS VHZ sample which shows any clear reported evidence of an AGN or LRD. COSMOS-52185, COSMOS-300001 and EGS-25297 \citep{donnan2025b} are three $z\simeq10$ galaxies with particularly extreme UV-continuum slope. The COSMOS pair are a close pair of potentially merging galaxies with extremely blue UV slopes. EGS-25297 has a very red slope potentially suggesting significant dust attenuation ($A_{\mathrm{V}}\simeq0.9$), which would imply a significantly high intrinsic UV luminosity $\sim2-3\times$ greater than the most luminous known $z>10$ galaxy to date, GN-z11 \citep[e.g.][]{bunker2023b}. 

\begin{figure*}[ht]
\centering
\includegraphics[width=\linewidth]
{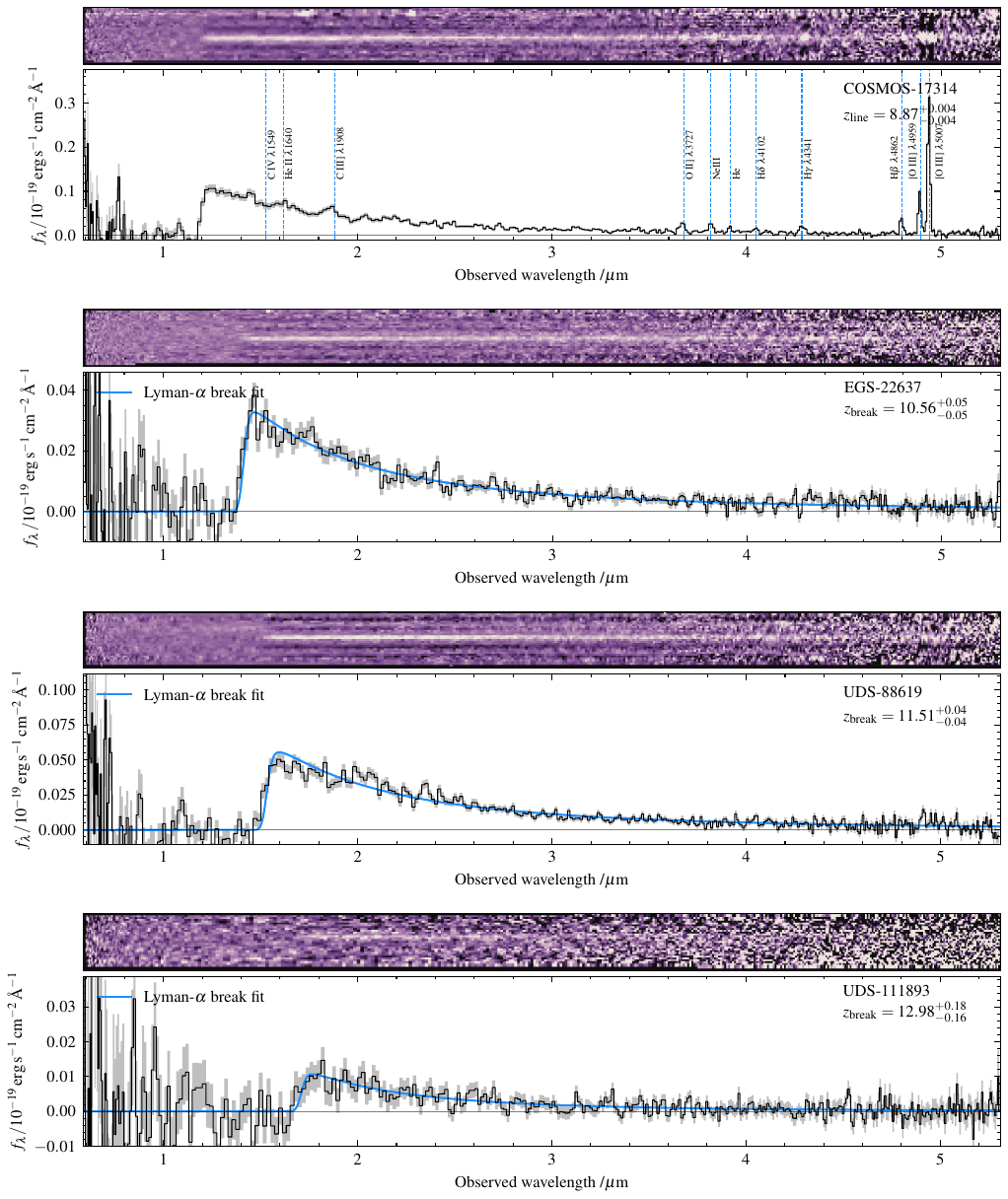}
\caption{Four example spectra of galaxies confirmed at $z_{\mathrm{spec}}>8.5$ from the CAPERS program at $z\sim9$, $z\sim10$, $z\sim11$ and $z\sim13$. For each spectrum we show the 2-D spectrum (top) and the extracted 1-D spectrum (bottom). Two of these sources are newly reported in this work.}
\label{fig:example_spectra}
\end{figure*}

There are a number of galaxies spectroscopically confirmed by the CAPERS program which are newly presented in this work. We show the NIRSpec PRISM spectra of four example sources (two of which are newly reported in this work) in Fig.~\ref{fig:example_spectra}.
This includes UDS-111893 which is the highest redshift source in CAPERS at $z_{\mathrm{spec}}=12.98$ and is the 7th highest redshift source confirmed to date. 

There are also some CAPERS spectra reported to have $z_{\mathrm{spec}}>8.5$ in the literature which we do not recover in this work. Firstly, \citet{robertsborsani2026} report EGS-65480 at $z_{\mathrm{spec}}=12.344$. However, we find the redshift inconclusive for this source due to the weak SNR and heavily damped shape of the Lyman-$\alpha$ break. If this break is indeed a Lyman-$\alpha$ break at high redshift, then we require an extremely high local hydrogen column density of $\log(N_{\mathrm{HI}}/ \mathrm{cm^2})\simeq24$. However, this break is also fully consistent with a Balmer break at $z\simeq3.1$. \citet{robertsborsani2026} also report EGS-43539 at $z_{\mathrm{spec}}=10.816$ which we also find inconclusive. This is once again partly due to a very damped Lyman-$\alpha$ break but also the NIRSpec data is absent at $\lambda < 1.4\, \mu \mathrm{m}$ due to a chip gap making it impossible to determine if there is the presence of any flux blueward of the observed break.

\section{Conclusion}
\label{sec:conclusion}

In this work we have used the large sample ($N=183$) of $z\geq8.5$ photometric galaxy candidates observed with the CAPERS spectroscopic program, to quantify the effectiveness of different methods of successfully selecting these sources. The conclusions are summarized below.

\begin{itemize}
    \item Out of the 183 CAPERS VHZ candidates with a successful spectroscopic redshift, we spectroscopically confirm 48 to be at $z_{\mathrm{spec}}\geq8.5$ through modeling of emission lines where available or the Lyman-$\alpha$ break when no lines are detected (8 sources).

    \item We test six commonly used spectral templates in \textsc{eazy} finding the highest success and completeness rates with template sets that include sufficiently blue continua to model galaxies at $z\geq8.5$ successfully.

    \item The success of high-redshift spectroscopic confirmation is strongly correlated with the signal-to-noise in the detection filter, the strength of the Lyman-$\alpha$ break and the $\Delta \chi^2$ between the low and high redshift solutions. In particular, a strong $\Delta \chi^2$ requirement is more effective at maintaining completeness while successfully removing low-redshift contaminants than simply strong SNR requirements.

    \item The use of a UV LF prior on the posterior redshift distribution from SED fitting is highly efficient at selecting high-redshift galaxies while maintaining high completeness outperforming even the most conservative SNR and $\Delta \chi^2$ requirements. 

    \item A number of VHZ candidates that are formally considered redshift failures ($|dz|>0.15$) are still confirmed at $z\gtrsim8$. These are cases where photometric redshift fitting interprets the strong flux in F410M and F444W incorrectly as a Balmer break rather than strong \oiiiboth\ and H$\beta$ emission.

    \item We report the observational properties of the 48 CAPERS sources confirmed at $z\geq8.5$ which reveal a systematic bias in photometric UV-slope measurements towards moderately bluer UV slopes due to an average overestimation of photometric redshifts.
\end{itemize}

Overall, this work quantifies a number of different selection criteria with the most efficient being the use of a UV LF prior on the derived $p(z)$ and therefore we recommend this for future selection of galaxies at $z\geq8.5$. The 48 confirmed sources at $z\geq8.5$ in CAPERS add to an ever growing sample of spectroscopic data in the early Universe.

\begin{acknowledgments}
This work is based on observations made with the NASA/ESA/CSA \textit{James Webb Space Telescope}, obtained at the Space Telescope Science Institute, which is operated by the Association of Universities for Research in Astronomy, Incorporated, under NASA contract NAS5-03127. Support for program number GO-6368 was provided through a grant from the STScI under NASA contract NAS5-03127. The data were obtained from the Mikulski Archive for Space Telescopes (MAST) at the Space Telescope Science Institute. 
These observations are associated with program \#6368, and can be accessed via \dataset[doi: 10.17909/0q3p-sp24]\\
We thank Duncan Austin, Rychard Bouwens, Maximilien Franco, and Haojing Yan for contributing high redshift candidates that were incorporated into CAPERS target prioritization.
We thank Eduardo Ba\~{n}ados for permission to use NIRCam F090W data from GO-2234 for the CEERS photometric catalog.
PAH acknowledges support from NASA under award 80GSFC24M0006.
\end{acknowledgments}




%
\facilities{JWST, HST}
\software{astropy \citep{2013A&A...558A..33A,2018AJ....156..123A}, \textsc{eazy} \citep{brammer2008}, 
          \textsc{bagpipes} \citep{carnall2018}, \textsc{dynesty} \citep{speagle2020}
          }


\appendix

\section{Lyman-$\alpha$ break modeling for redshift measurement}
\label{sec:apdx_z_break}
In Section~\ref{sec:z_spec_measurement} we introduce a Lyman-$\alpha$ break redshift fitting code to measure a spectroscopic redshift for NIRSpec PRISM spectra that are absent of emission lines. In this appendix we describe this code in further detail and motivate the use of the model priors used in this code, in particular testing the need for additional damping through a local neutral hydrogen reservoir. 

To test the optimum model and priors for obtaining accurate spectroscopic redshifts, we first construct a sample of CAPERS sources at $6.5<z_{\mathrm{spec}}<9.5$ for which we have an emission line redshift. This sample contains 92 sources with complete wavelength coverage and sufficient SNR at the Lyman-$\alpha$ break ($\mathrm{SNR_{1300 \AA}/pix} >2$). We then test the accuracy of recovering the known emission line spectroscopic redshift through fitting of the Lyman-$\alpha$ break. 

\begin{figure*}[ht]
\centering
\includegraphics[width=\linewidth]
{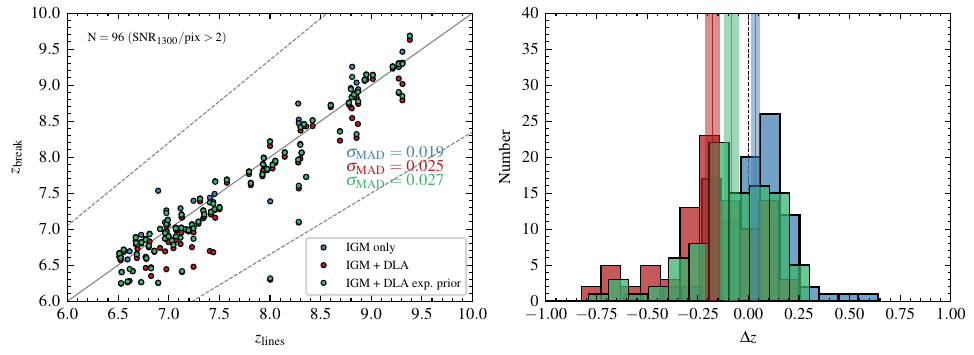}
\caption{\textit{Left}: A comparison of the Lyman-$\alpha$ break redshift to the emission line redshift for 96 CAPERS sources with $\mathrm{SNR_{1300 \AA}/pix} >2$ at $6.5<z_{\mathrm{spec}}<9.5$ where damping is driven by IGM only (blue), IGM and a local hydrogen reservoir (damped Ly$\alpha$ absorbers, DLAs) with a log-uniform prior (red) and where the prior is given by an exponential (green). The dashed lines show where $dz=\pm0.15$, where $dz = (z_{\mathrm{phot}} - z_{\mathrm{spec}})/(1+z_{\mathrm{spec}})$. Inset text notes the scatter given by $\sigma_{\mathrm{MAD}} = 1.4826\, \mathrm{MAD}$, where $\mathrm{MAD}$ is the median absolute deviation of $dz$. \textit{Right}: The distribution of $\Delta z$ for each Lyman-$\alpha$ break-fit configuration with their respective median value indicated by the solid line. The shaded region shows the standard error on the median.} 
\label{fig:break_test_zz}
\end{figure*}

Our initial test uses a model with four free parameters which are the redshift, $z$, UV-slope, $\beta$, the IGM neutral fraction, $X_{\mathrm{HI}}$ and the normalization, $A$. These have the following uniform priors: $6.5\leq z\leq20$, $-4\leq \beta\leq4$, $0\leq X_{\mathrm{HI}}\leq1$ and $0.0001\leq A \leq0.5$. We use the nested sampling code \textsc{dynesty} \citep{speagle2020} to perform this fitting. A comparison of the resultant Lyman-$\alpha$ break redshift and the emission line redshift for this test sample is shown by the blue data-points in Fig.~\ref{fig:break_test_zz}.

This shows generally good accuracy but with a slight average overestimation of the spectroscopic redshift of $\Delta z \simeq 0.03$. A possible cause is a failure to allow sufficient damping of the Lyman-$\alpha$ break in our model which motivates the inclusion of a further free parameter which parameterizes the column density of a local hydrogen reservoir \citep{heintz2023} with the following uniform prior: $18\leq \log(N_{\mathrm{HI}}/\mathrm{cm}^2) \leq24$. The results of this fitting is shown by the red data-points in Fig.~\ref{fig:break_test_zz}. This now on average causes a significant underestimation in spectroscopic redshift of $\Delta z \simeq 0.18$ with increased scatter suggesting that for a number of sources, the Lyman-$\alpha$ break is being modeled a local column density too high and therefore worsening the average best-fitting Lyman-$\alpha$ break redshift.

Finally, we test a non-uniform prior heavily weighted to low column densities to investigate if this produces an improved fit where the prior is given by $p(\log(N_{\mathrm{HI}}/\mathrm{cm}^2)) \propto e^{18-\log(N_{\mathrm{HI}}/\mathrm{cm}^2)}$. These results are shown by the green data-points in Fig.~\ref{fig:break_test_zz}. This provides an improvement over a uniform prior with the redshift now underestimated by $\Delta z \simeq 0.08$. However, in comparison to the IGM only model there is still greater redshift inaccuracy on average and increased scatter (likely due to the increased number of free parameters). Therefore, for our fiducial Lyman-$\alpha$ break redshift measurements, we adopt the IGM only model in this work.


\section{CAPERS-UDS-z13}
\label{sec:apdx_uds-z13}

\begin{figure*}[b]
\centering
\includegraphics[width=\linewidth]
{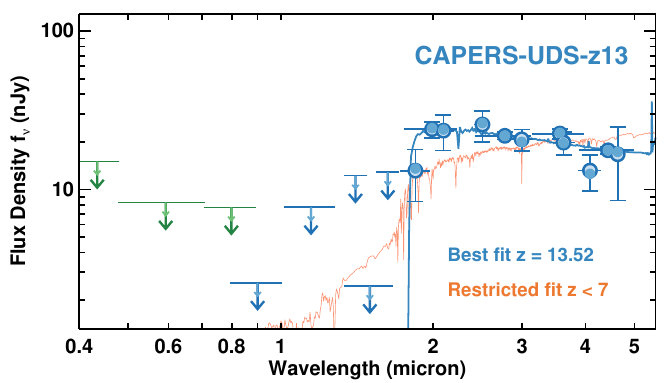}
\caption{Photometric spectral energy distribution for CAPERS-UDS-z13 including medium-band photometry from MINERVA \citep{muzzin2025}, overlaid with photometric redshift spectral templates.  The blue model shows the best-fitting template at $z_{\mathrm{phot}} = 13.52$, while the red model shows the best-fitting low-redshift solution ($z_{\mathrm{phot}} = 3.56$. The large $\Delta\chi^2 = 46.74$ between these models strongly favors the high-redshift solution.  Deep NIRCam F090W and F150W images set tight upper limits on flux below the photometric break, strongly excluding low-redshift solutions.}
\label{fig:sed_uds_z13}
\end{figure*}

As noted in \S\ref{sec:vhz_selection}, the highest secure spectroscopic redshift measured in CAPERS is $z = 12.98^{+0.18}_{-0.16}$ for UDS-111893, which we nickname CAPERS-UDS-z13.  This candidate was originally selected by four different VHZ subsamples (i.e., VHZ count = 4 as per \S\ref{sec:vhz_selection}), including the robust ``UVHZ'' selection \citep[as in][]{finkelstein2024}. It was included in CAPERS MSA plans for UDS pointing P6, originally scheduled for observation on UT 10 January 2025, but those visits and others were not executed due to the evacuation of the Jet Propulsion Laboratory during the extremely destructive wildfire centered in Altadena, CA. That shutdown led to limitations on data transfer from JWST via the Deep Space Network. The UDS-P6 observations were rescheduled for January 2026. In the meanwhile, the JWST MINERVA program \citep[GO-7814,][]{muzzin2025} obtained its first epoch of NIRCam imaging through eight medium-band filters, covering about half the PRIMER-UDS field, including CAPERS-UDS-z13.\footnote{The MINERVA team's public data products are available at \url{https://minerva.colorado.edu}.} We reduced the MINERVA data (H.\ Akins, private communication), as well as other public NIRCam data taken since the original PRIMER observations, producing new UDS image mosaics and new UNICORN (\S\ref{sec:photometry}) photometric catalogs. With the additional photometric information (Fig.~\ref{fig:sed_uds_z13}), CAPERS-UDS-z13 was again assigned high priority for spectroscopy, and the CAPERS NIRSpec observations were successfully executed on UT 12 January 2026, with a total integration time of 17,069s.

The spectrum of CAPERS-UDS-z13 is shown in the bottom panel of Fig.~\ref{fig:example_spectra}. Its only clear feature is a break at about $1.7\,\mu$m. Deep images with the NIRCam F090W and F150W filters \citep[21452s and 15560s, respectively, from BEACON, GO-3990,][]{morishita2025} set tight upper limits on flux below the break, yielding $m({\mathrm{F090W,F150W}}) < 30.3$ at $2\sigma$.  These limits contribute to a very large $\Delta \chi^2 = 46.7$ between low and high photometric redshift solutions, ruling out a Balmer Break interpretation. No convincing emission lines are detected in the spectrum. Therefore, we derive the spectroscopic redshift using the Lyman-$\alpha$ break fitting methodology described in Appendix~\ref{sec:apdx_z_break}.  The observed break is somewhat ``soft'', suggesting the presence of significantly damped Lyman-$\alpha$ absorption. A local DLA component in addition to the IGM absorption is not necessary to adequately model the break, but if included (as per Appendix~\ref{sec:apdx_z_break}) the redshift fit would be reduced and broadened to $z = 12.78^{+0.52}_{-0.26}$.

The spectral SNR is disappointingly low for a galaxy with $m({\mathrm{F200W}}) = 27.8$, most likely due to the fact that the target was positioned near one corner of the open micro-shutter aperture, leading to significant pathloss.  As a result, upper limits on UV emission lines are not very tight, particularly given the uncertainty in the redshift determined from the break. Following the method from \citet{donnan2026}, we obtain $2\sigma$ upper limits on the rest-frame equivalent widths of \niv\,, \civ\, and \ciii\ of $\simeq47\, \mathrm{\AA}$. This limit is similar to the strengths of the strongest measured rest-frame UV lines in galaxies at $z>10$ \citep[e.g. GHZ2 with $\mathrm{EW_{CIV,0}}\simeq46\, \mathrm{\AA}$;][]{castellano2024}. Nevertheless, it is notable that the galaxy is significantly extended in the deep F444W image (also from BEACON).  Several previous studies have noted an apparent correlation between compact/extended morphology and strong/weak UV line emission \citep[e.g.,][]{harikane2025,robertsborsani2026,donnan2026}.

CAPERS-UDS-z13 has an absolute UV magnitude $M_{\mathrm{UV}} = -20.03$, less luminous than the brightest known galaxies at $12 < z < 14.5$ (see Figure~\ref{fig:Muv_z}), and a moderately blue UV spectral slope $\beta = -2.42^{+0.34}_{-0.35}$. We also perform spectro-photometric modeling of the NIRSpec and NIRCam data simultaneously using \textsc{bagpipes} \citep{carnall2018}, following the method used in \citet{donnan2026}. We use the ``bursty continuity'' non-parametric prior \citep{tacchella2022} on the SFH and obtain a stellar-mass of $\log(M_{\star}/ \mathrm{M_{\odot}})=8.3^{+0.7}_{-0.3}$ and minimal dust attenuation of $A_{\mathrm{V}}=0.04^{+0.04}_{-0.03}$, consistent with the blue spectral slope.


\bibliography{vhz}{}
\bibliographystyle{aasjournalv7}



\end{document}